\documentclass[a4paper,twocolumn,nofootinbib,superscriptaddress]{revtex4-1}

\usepackage[pdftex]{graphicx}
\usepackage{amsthm}
\usepackage{bbold}
\usepackage{braket}
\usepackage{setspace}
\usepackage{enumerate}
\usepackage{amsmath,amssymb,amsfonts}
\usepackage[utf8]{inputenc}
\usepackage[T1]{fontenc}
\usepackage{fancyhdr, lastpage}
\usepackage[toc,page]{appendix}
\usepackage[caption=false]{subfig}
\usepackage{hyperref}
\usepackage{mathenv}
\usepackage[top=1.2in, bottom=1.2in, left=0.5in, right=0.5in]{geometry}

\newcommand{\del}[1]{\nabla}

\newcommand{\om}{\omega}

\newcommand{\average}[1]{\langle{#1} \rangle}

\renewcommand{\del}{\nabla}

\newcommand{\be}{\begin{equation}}
\newcommand{\ee}{\end{equation}}

\begin{document}
\title{Chiral quantum optics with V-level atoms and coherent quantum feedback}
\date{\today}

\author{Pierre-Olivier Guimond}
\affiliation{Institute for Quantum Optics and Quantum Information of the Austrian Academy of Sciences, A-6020 Innsbruck, Austria}
\affiliation{Institute for Theoretical Physics, University of Innsbruck, A-6020, Innsbruck, Austria}
\author{Hannes Pichler}
\affiliation{Institute for Quantum Optics and Quantum Information of the Austrian Academy of Sciences, A-6020 Innsbruck, Austria}
\affiliation{ITAMP, Harvard-Smithsonian Center for Astrophysics, 60 Garden Street, Cambridge, MA 02138, USA}
\affiliation{Physics Department, Harvard University, 17 Oxford Street, Cambridge, Massachusetts 02138, USA}
\author{Arno Rauschenbeutel}
\affiliation{Vienna Center for Quantum Science and Technology, Atominstitut, Vienna University of Technology, 1020 Vienna, Austria}
\author{Peter Zoller}
\affiliation{Institute for Quantum Optics and Quantum Information of the Austrian Academy of Sciences, A-6020 Innsbruck, Austria}
\affiliation{Institute for Theoretical Physics, University of Innsbruck, A-6020, Innsbruck, Austria}

\begin{abstract}
 We study the dissipative dynamics of an atom in a V-level configuration driven by lasers and coupled to a semi-infinite waveguide. The coupling to the waveguide is chiral, in that each transition interacts only with the modes propagating in a given direction, and this direction is opposite for the two transitions. The waveguide is terminated by a mirror which coherently feeds the photon stream emitted by one transition back to the atom. First, we are interested in the dynamics of the atom in the Markovian limit where the time-delay in the feedback is negligible. Specifically, we study the conditions under which the atom evolves towards a pure "dark" stationary state, where the photons emitted by both transitions interfere destructively thanks to the coherent feedback, and the overall emission vanishes. This is a single-atom analog of the quantum dimer, where a pair of laser-driven two-level atoms is coupled to a unidirectional waveguide and dissipates towards a pure entangled dark state. Our setup should be feasible with current state-of-the-art experiments. Second, we extend our study to non-Markovian regimes and investigate the effect of the feedback retardation on the steady-state.
 \end{abstract}
\maketitle
\section{Introduction}

The ability to engineer the coupling between quantum optical systems and photonic baths allows for many applications in quantum information \cite{quantuminternet,Reiserer:2015en}, such as the preparation of single- or many-body quantum states via the dissipative emission of photons \cite{dissip1,dissip2,dissip3}. During the last decade, tremendous experimental progress has been made to efficiently couple atoms (either real or artificial) to one-dimensional waveguides \cite{Hoi:2015fh,vanLoo:2013df,Astafiev:2010cm,Lodahl:2015fy,Riedinger:2016cl,Fang:2016ka}, for instance with real atoms coupled to optical fibers \cite{PhysRevLett.110.243603,PhysRevLett.113.143601} or photonic structures \cite{nat1,Thompson1202,Tiecke:2014aa}. For the last few years there has been a strong interest towards the implementation of \emph{chiral} couplings between atoms and waveguides, where by "chiral" we mean that the coupling depends on the propagation direction of the photons in the waveguide. Several recent experiments have demonstrated such chiral couplings between quantum emitters and guided light fields, for instance using atoms coupled to the evanescent field of whispering-gallery-modes bottle microresonators \cite{PhysRevLett.110.213604} and tapered optical fibers \cite{Mitsch:2014aa}, or quantum dots in photonic nanostructures \cite{PhysRevLett.110.037402,Sollner:2015aa}.

\begin{figure}[h!]
\includegraphics[width=9cm]{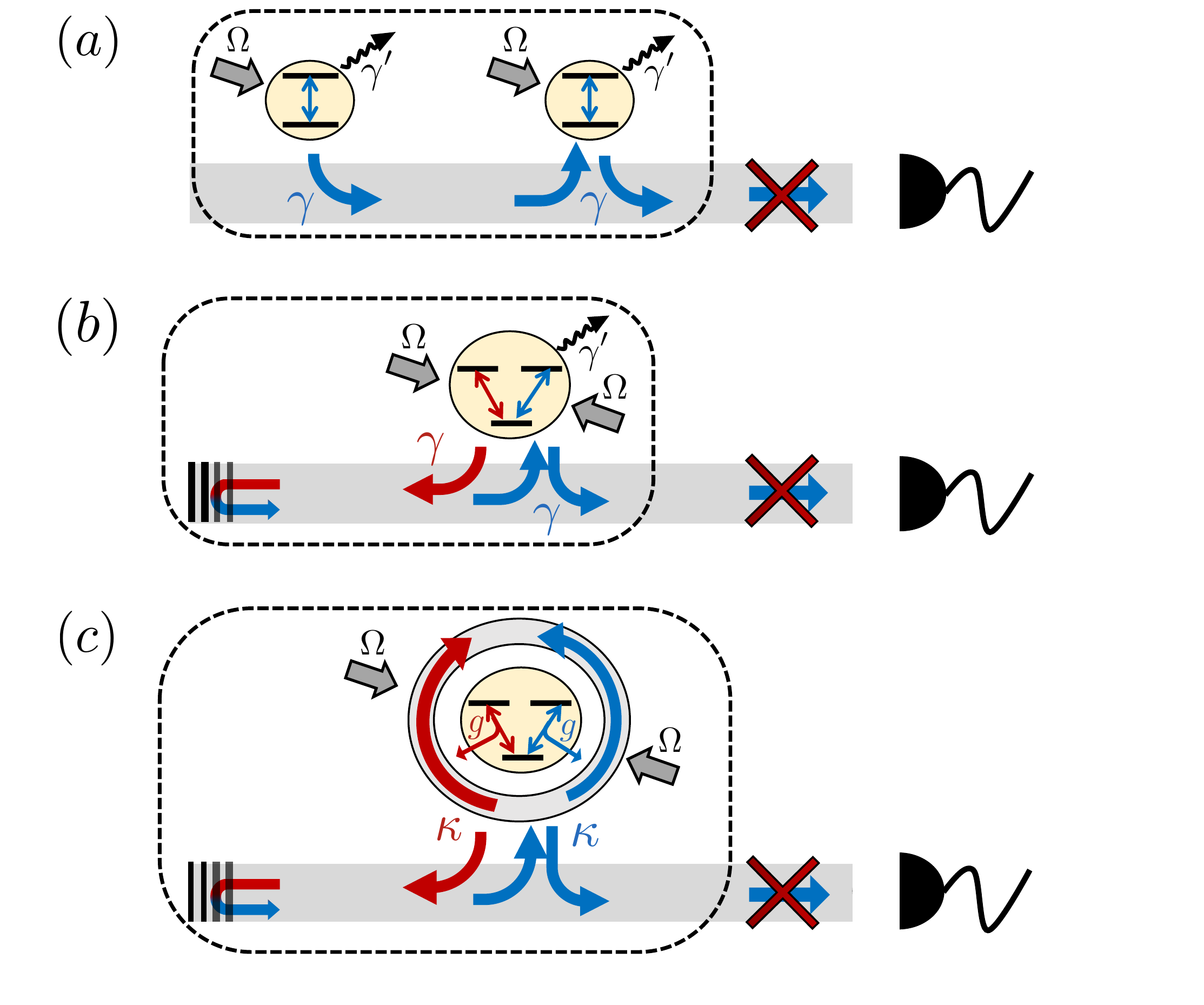}
\caption{ \label{fig:vsyswithwaveguide}(Color online) (a) A pair of two-level systems (TLS) coupled to a unidirectional waveguide. The TLSs are driven by classical fields with Rabi frequencies $\Omega$ and are additionally coupled to the guided modes with a rate $\gamma$ and to non-guided modes with a rate $\gamma'$ which we will first assume to be negligible. Under specific conditions, the photons emitted by the left TLS are fully coherently absorbed by the right one. Remarkably, the whole system then relaxes towards a pure \emph{dark state}, where the photon emission vanishes \cite{1367-2630-14-6-063014}. See Sec.\,\ref{sec:dimers}. 
(b) Atom in a V-level configuration driven with a Rabi frequency $\Omega$ and coupled to a bidirectional waveguide, where the two transitions couple to the guided modes propagating in opposite directions, represented with red and blue arrows, with the same rate $\gamma$. The photon stream propagating to the left is fed back to the system, which now acts as its own coherent absorber. Under the appropriate conditions, the system relaxes towards a pure dark state, analogously to the case described in (a). See Sec.\,\ref{subsect:markovmasterequation} and Sec.\,\ref{subsec:darkstates}. (c) The cooperativity can be increased by coupling the atom with a rate $g$ to a cavity which preserves chiral coupling \cite{PhysRevLett.110.213604} and is also coupled with a rate $\kappa$ to the guided modes. In the bad-cavity regime $g\ll\kappa$, the system reproduces the physics described in (b), with an effective atom-waveguide coupling of $\gamma\approx(2g)^2/\kappa$ and a negligible decoherence rate $\gamma'$. See Sec.\,\ref{sec:couplcavi}.}
\end{figure}

Chiral couplings allow for the formation of entangled states as \emph{pure} steady-states of the dynamics of laser-driven open systems via the dissipative emission of photons \cite{PhysRevLett.113.237203,PhysRevA.91.042116,ramos2016}. In particular, it was shown in Ref.\,\cite{1367-2630-14-6-063014} that an ensemble of atomic two-level systems (TLS) driven by classical fields and coupled to a unidirectional waveguide eventually stops emitting photons in the waveguide under the right conditions on the driving fields. The atoms then form EPR-correlated pairs (see Fig.\,\ref{fig:vsyswithwaveguide}(a)) where the photons emitted by the first atom are coherently absorbed by the second one. The atomic pair, called quantum \emph{dimer}, now forms a pure \emph{dark} state $\rho(t)\to\ket{D}\bra{D}$ which is decoupled from the waveguide. This steady-state is of the form $\ket{D}\propto \ket{gg}+\alpha (\ket{eg}-\ket{ge})$, where $\ket{g}$ and $\ket{e}$ are the ground and excited states of each TLS and $\alpha$ depends on the setup parameters. Only the excited components of the state contribute to the photon radiation in the open waveguide, and their total contribution vanishes thanks to their opposite sign.

While the long-term goal is to observe this phenomenon in the laboratory, the purpose of the present work is to explore the possibility of experimenting analogous physics with a single atom, which would be achievable at the current state of the technology. The system of interest is represented in Fig.\,\ref{fig:vsyswithwaveguide}(b), and is constituted of a single atom coherently driven by its quantum feedback and by external lasers. The atom has a V-level configuration where the transitions are coupled to the guided modes propagating in opposite directions. The first question we want to address is thus whether the dynamics of the two atoms of Fig.\,\ref{fig:vsyswithwaveguide}(a) can be mimicked by the interaction of one atom with its mirror image. The pumping to a dark state in both setups can be achieved only if the coupling to external non-guided modes $\gamma'$ is negligible. Recent experiments coupling atoms to waveguides (albeit without featuring chiral couplings) have reported $\beta$-factors of $\beta\equiv\gamma/(\gamma+\gamma')\approx 0.5$ \cite{PhysRevLett.115.063601,hood2016}, with $\gamma$ the coupling to the guided modes. On the other hand, very high $\beta$-factors of $\beta=0.98$ have been reached using quantum dots as artificial atoms \cite{PhysRevLett.113.093603}. As an alternative way to increase the coupling strength between the atom and the guided modes, we consider the setup of Fig.\,\ref{fig:vsyswithwaveguide}(c), where the atom is coupled with a rate $g$ to a cavity which preserves chiral coupling \cite{PhysRevLett.110.213604} and is itself strongly coupled to the guided modes with a rate $\kappa$. In the bad-cavity regime $g\ll\kappa$, the atom undergoes the same dynamics as in Fig.\,\ref{fig:vsyswithwaveguide}(b), with $\gamma\approx(2g)^2/\kappa$ and an increased $\beta$-factor \cite{Tiecke:2014aa,PhysRevX.5.041036}.

Moreover, our system is a very simple example of coherent quantum feedback \cite{Wiseman:2010vw}, and from a theory view point the effect of the delay in the feedback has recently attracted a lot of interest \cite{PhysRevLett.115.060402,PhysRevLett.116.093601,Nemet:2016wg,AlvarezRodriguez:2016vh,SanchezBurillo:2016uz,PhysRevLett.113.183601}. In general, one of the requirements for the existence of pure atomic states is that the photon number between the mirror and the atom is negligible, otherwise the atom would become entangled with these photons. This assumption is equivalent to the usual \emph{Markovian} approximation, where the retardation effects in the effective atomic dynamics, induced by the finite photon travel time, are neglected in order to derive an atomic master equation \cite{1367-2630-14-6-063003, PhysRevA.91.042116,ramos2016}. The second question we want to address is thus how this finite delay affects the properties of the steady-state. In order to do so, an approach has been developed in Ref.\,\cite{PhysRevLett.116.093601}, which employs matrix-product states (MPS) techniques \cite{RevModPhys.77.259} to track the entangled state of the atom and of the photons, and dynamically solves the quantum stochastic Schr\"odinger equation \cite{gardinerzoller_quantumnoise} (QSSE).

The paper is organised as follows. In Sec.\,\ref{sec:dimers}, we briefly review the physics of the quantum dimer formation with two-level atoms. In Sec.\,\ref{sec:markov} we address the dynamics of our feedback system in the Markovian limit. We derive the master equation and analyze the conditions under which the system dissipates towards a pure dark state. We also show that by coupling the atom to the cavity of Fig.\,\ref{fig:vsyswithwaveguide}(c) one can increase the $\beta$-factor. In Sec.\,\ref{section:nonmarkov}, we investigate how the steady-state properties are modified by the retardation effects of a non-Markovian coherent feedback. Finally, in Sec.\,\ref{sec:experiment} we discuss some experimental considerations such as the effect of the coupling to external non-guided modes and the effect of an imperfect chiral coupling to the waveguide. 

\section{Dimerization of an atomic chain}
\label{sec:dimers}

To provide the basis for the understanding of the physics of our feedback system (Fig.\,\ref{fig:vsyswithwaveguide}(b)), we first review the formation of entanglement in \emph{cascaded} many-body two-level atoms \cite{PhysRevLett.70.2273,PhysRevLett.70.2269}, i.e in an ensemble of atoms coupled to a unidirectional waveguide \cite{gardinerzoller_quantumworld}. 

Let us first consider two atoms driven by classical fields near resonance and coupled with a rate $\gamma$ to the guided modes, as represented in Fig.\,\ref{fig:vsyswithwaveguide}(a). The atoms are separated by a distance $d$ along the waveguide. We consider the ideal case where the coupling to the guided modes is perfectly chiral and the coupling to non-guided modes $\gamma'$ is negligible. Due to the unidirectionality of the waveguide, atom $1$, on the left, does not feel the presence of atom $2$, on the right. The second atom however is continuously driven by the coherent photon emission of the first one. If we neglect the travel time of the photons between both atoms (Markovian approximation), we can derive a master equation for the atoms, which reads (with $\hbar=1$) \cite{1367-2630-14-6-063014} \begin{equation}\label{eq:dimermasterequation} \frac{d\rho}{dt}=-i[H_S,\rho]+2\gamma\mathcal D[\sigma_\text{tot}^-]\rho, \end{equation}
where $\mathcal D$ is the Lindblad superoperator  \be \mathcal D[a](\cdot) \equiv a\cdot a^\dagger -\frac12\{a^\dagger a,\cdot\}.\ee
The Hamiltonian reads \begin{equation}\begin{aligned} \label{eq:HSdimers}H_S = & -\big(\delta_1 \ket{e_1}\bra{e_1}+\delta_2 \ket{e_2}\bra{e_2}\big) \\ &-\frac{\Omega}2\big(\sigma_1^- + e^{i\phi'}\sigma_2^-+\text{H.C.}\big)\\ &+ i\frac{\gamma}2\big(e^{i\phi}\sigma_1^+\sigma_2^--e^{-i\phi}\sigma_2^+\sigma_1^-\big),\end{aligned}\end{equation} where $\ket{g_i}$ and $\ket{e_i}$ are the ground and excited states of atom $i=1,2$, and $\sigma_i^- =\ket{g_i}\bra{e_i}$. The first two rows of Eq.\,\eqref{eq:HSdimers} are the laser driving terms, with $\delta_i = \bar\omega-\omega_i\ll\omega_i$ the detuning between the laser frequency $\bar\omega$ and the transition frequency for atom $i$. The drivings of the two atoms have the same Rabi frequency $\Omega$ and a relative phase $\phi'$. The last row is the dipole-dipole interaction induced by the exchange of photons through the waveguide. The phase acquired by a photon traveling between both atoms is $\phi=-\bar\omega  d/c$ where $c$ is the speed of light in the waveguide. The unidirectionality of the problem lies in this dipole-dipole term, which is asymmetric under the exchange of the labels $1$ and $2$. The last term of Eq.\,\eqref{eq:dimermasterequation} is the collective decay of the ensemble at a {superradiant rate} $2\gamma$, where the collective jump operator is \begin{equation}\label{eq:sigmatot}\sigma_\text{tot}^- = \big(\sigma_1^-+e^{i\phi}\sigma_2^-\big)/\sqrt{2}. \end{equation}

Although two phases appear in Eq.\,\eqref{eq:HSdimers}, only their difference $\phi-\phi'$ affects the dynamics. We can thus gauge $\phi'$ away in Eq.\,\eqref{eq:HSdimers} and Eq.\,\eqref{eq:sigmatot} by redefining the state $\ket{e_2}\to e^{i\phi'}\ket{e_2}$ and the phase $\phi\to\Delta\phi=\phi-\phi'$. We now make the assumption that $\Delta\phi$ is a multiple of $2\pi$, which we will refer to as the \emph{commensurability condition}. Let us consider two scenarios. In the first scenario, the atoms are driven by different lasers outside of the waveguide. The relative laser phase $\phi'$ can be independently tuned, and should be set to $\phi$ in order to satisfy the commensurability. In the second scenario, the atoms are driven by the same laser propagating \emph{inside} the fiber. In that case, $\phi'$ is no longer an independent parameter and is equal to $\phi$ by definition. Here, the commensurability is thus automatically satisfied.

We are interested in the formation of pure atomic steady-states $\rho(t)\to\ket{D_{12}}\bra{D_{12}}$, which have to be in particular disentangled from the waveguide. This requires that, for these states, the system should effectively decouple from the waveguide and stop radiating photons, hence their appellation of "dark" states. In terms of Eq.\,\eqref{eq:dimermasterequation}, this means that the jump operator $\sigma_\text{tot}^-$, which induces decoherence, should annihilate the steady-state. This restricts us to states of the form \begin{equation}\label{eq:dimerD}\ket{D_{12}}=\frac{1}{\sqrt{1+|\alpha|^2}}\big(\ket{g_1g_2}+\alpha \ket{S_{12}}\big),\end{equation}
where $\ket{S_{12}}=(\ket{e_1g_2}-\ket{g_1e_2})/\sqrt{2}$ is the "singlet" state. In order to be a steady-state of the Hamiltonian $H_S$, one can show that the detunings have to be opposite ($\delta_1=-\delta_2$), which yields \begin{equation}\label{eq:alphadimer} \alpha= -\sqrt{2}\frac{\Omega}{i\gamma + 2\delta_1}. \end{equation}

\begin{figure}
\includegraphics[width=9cm]{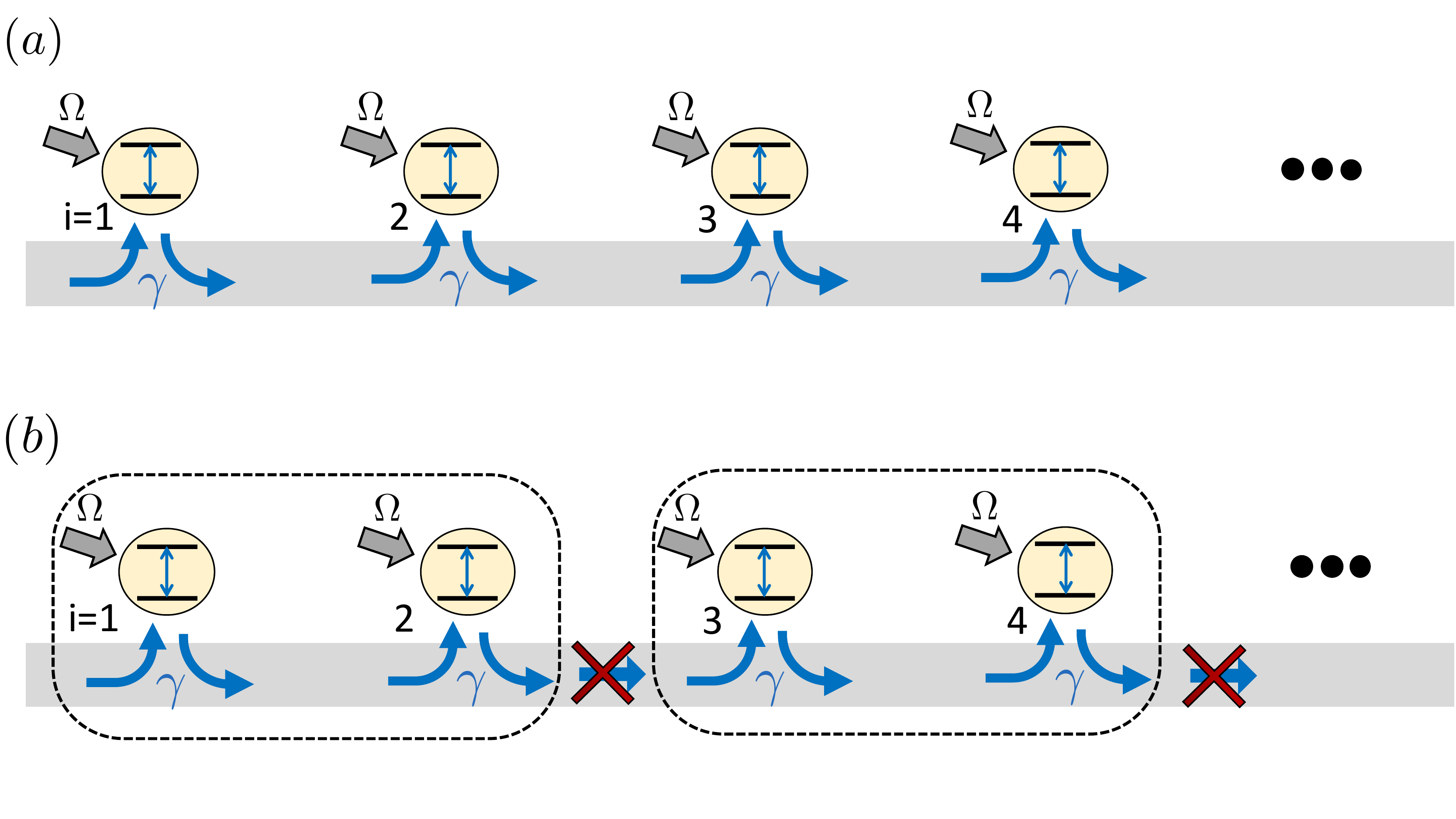}
\caption{\label{presentation3}(Color online) (a) Ensemble of laser-driven two-level atoms coupled to a unidirectional waveguide. (b) As time evolves, the atoms form pairs. These quantum dimers are in a pure dark state and do not exchange or radiate photons.}
\end{figure}

Remarkably, this result can be easily generalized to the situation represented in Fig.\,\ref{presentation3}(a), where an array of $2N$ atoms is coupled to the waveguide. Provided $\delta_1=-\delta_2$, the first two atoms will be pumped to the dark state $\ket{D_{12}}$ since they are not influenced by the presence of the other atoms thanks to the directionality of the problem. As their state converges towards this steady-state, the photon stream between atoms 2 and 3 vanishes, and finally the third and fourth atoms do not feel the presence of the first two ones either, which allows them to evolve towards their dark state $\ket{D_{34}}$, provided $\delta_3=-\delta_4$.
By iterating this argument, the steady-state of the atomic array will factorize into \be \ket{D}=\ket{D_{12}}\otimes\ket{D_{34}}\otimes...\ee This is represented in Fig.\,\ref{presentation3}(b). Each atomic pair in a dark state is called quantum \emph{dimer}, and the factorization process is the dimerization of the atomic chain.

\section{Markovian quantum feedback}
\label{sec:markov}

In this section we analyze the dynamics of the atom of our coherent feedback system (Fig.\,\ref{fig:vsyswithwaveguide}(b)) in the Markovian limit where the feedback retardation effects are neglected, and we expose the similarities and the differences with the quantum dimers of the previous section. In Sec.\,\ref{subsect:markovmasterequation}, we derive the master equation for the atom in the ideal case where the coupling to non-guided modes is negligible. In Sec.\,\ref{subsec:darkstates}, we study the dissipative evolution of the system towards the steady-state and derive the conditions under which this state is pure. In Sec.\,\ref{sec:couplcavi} we take the non-guided modes into account and we discuss the possibility of increasing the cooperativity in this setting by coupling the atom to a cavity in the bad-cavity regime.

\subsection{Master equation}
\label{subsect:markovmasterequation}
The system consists of a single laser-driven atom in a V-level configuration. The atom is coupled to a semi-infinite waveguide terminated by a mirror located at a distance $d$ from the atom, as represented in Fig.\,\ref{fig:vsyswithwaveguide}(b). We denote the atomic ground state by $\ket{g}$, and the excited states by $\ket{e_1},\ket{e_2}$. The transition operators are given by $\sigma_{i}^-=\ket g\bra{e_i} (i=1,2)$. Additionally, the $\sigma_1$ transition is exclusively coupled to the guided modes propagating towards the mirror whereas the $\sigma_2$ transition is coupled to the modes propagating outwards. We assume that the dispersion relation of the waveguide is approximately linear around the laser frequency $\bar\omega$ over a relevant bandwidth $\theta\ll\bar\om$ (i.e. $\omega\approx c|k|$ with $\omega$ the mode frequency, $k$ the wave number along the propagation axis and $c$ the speed of light in the waveguide). 

The Hamiltonian for the waveguide is thus given by (with $\hbar=1$)\be H_B = \int_{\bar\om-\theta}^{\bar\om+\theta}d\om\,\omega\,b^\dagger_\om b_\om, \ee where $b_\omega$ annihilates a photon with frequency $\omega$, and $[b_\omega,b^\dagger_{\omega'}]=\delta(\omega-\omega')$. The Hamiltonian for the driven atom reads \be\begin{aligned} H_a= \omega_1&\ket{e_1}\bra{e_1} + \omega_2\ket{e_2}\bra{e_2}\\-&\frac{\Omega}2\big(e^{i\bar\omega  t}\sigma_1^-+e^{i\bar\omega  t}e^{i\phi'}\sigma_2^-+\text{H.C.}\big).\end{aligned}\ee where $\omega_i$ are the transition frequencies, $\Omega$ is the Rabi frequency which we assume real and positive without loss of generality, and $\phi'$ is the relative driving phase. We used a rotating wave approximation (RWA), which is valid for $\Omega, |\delta_i|\ll\bar\omega $, where $\delta_i=\bar\omega -\omega_i$ is the detuning between the laser and the transition frequencies. The atom-waveguide interaction Hamiltonian in the RWA is given by  \be \label{eq:Hintdefdsa}H_\text{int} = i \int_{\bar\om-\theta}^{\bar\om+\theta}d\om\, g(\omega)b^\dagger_\omega\big(\sigma_1^- +\sigma_2^-e^{i(\pi-2\omega d/c)}\big)-\text{H.C.},\ee where $g(\omega)$ is the atom-waveguide coupling and the phase factor is the feedback photon phase which accounts for the propagation over a distance $2d$ and for the $\pi$-shift due to the mirror reflexion. Analogously to the case of Sec.\,\ref{sec:dimers}, the only physically relevant phase here is the phase difference $\pi-2\om d/c-\phi'$ between the feedback and the driving phases, hence we will gauge $\phi'$ away by redefining the state $\ket{e_2}\to e^{i\phi'}\ket{e_2}$. We will also assume that the coupling is approximately independent of the frequency over the relevant bandwidth $[\bar\om-\theta,\bar\om+\theta]$ and replace $g(\omega)\to\sqrt{\gamma/2\pi}$.

In order to derive the master equation for the atom, we move to an interaction picture with respect to the waveguide Hamiltonian $H_B$ and to a frame rotating with the laser frequency $\bar\omega $ for the atomic transitions. In this picture, the total Hamiltonian now reads $H_\text{tot}^I=H_{a}^I + H_\text{int}^I$, where 
\begin{equation}\label{eq:hsysdef1}\begin{aligned} H_a^I=-\delta_1\ket{e_1}\bra{e_1} - \delta_2\ket{e_2}\bra{e_2}-\frac{\Omega}2\big(\sigma_1^-+\sigma_2^-+\text{H.C.}\big).\end{aligned}\end{equation} 
and \begin{eqnarray} \nonumber H_\text{int}^I = i \sqrt{\frac{\gamma}{2\pi}}&& \int_{\bar\om-\theta}^{\bar\om+\theta}d\om\,b^\dagger_\omega \big(e^{-i(\bar\omega -\om)t}\sigma_1^- \\ &&+ e^{-i(\bar\om -\om)(t-\tau)}e^{i\Delta\phi}\sigma_2^-\big)-\text{H.C.},\label{eq:hintdef}\end{eqnarray} 
with $\tau=2d/c$ the time-delay of the quantum feedback and $\Delta\phi=\pi- \bar\om \tau-\phi'$ the phase difference which we will restrict to the interval $[-\pi,\pi]$ for convenience. From now on we will drop the superscript $I$ and always refer to the Hamiltonians in this picture. The state of the system comprising the atom and the waveguide at time $t$ $\ket{\Psi(t)}$ is related to the initial state $\ket{\Psi_0}$ by a unitary operator $U(t)$ such that $\ket{\Psi(t)}=U(t)\ket{\Psi_0}$, which satisfies the Schr\"odinger equation $i\frac{d}{dt}U(t)=H_\text{tot}U(t)$. In the Heisenberg picture, the waveguide operators $b_\om(t)=U^\dagger(t) b_\om U(t)$ satisfy the Heisenberg equation 
\be \label{heisenbergbom}\frac{d}{dt}b_\om(t) = \sqrt{\frac\gamma{2\pi}}\big(e^{-i(\bar\omega -\om)t}\sigma_1^-(t)+e^{-i(\bar\om-\om)(t-\tau)}e^{i\Delta\phi}\sigma_2^-(t)\big)\ee
where $\sigma_i^-(t)=U^\dagger(t) \sigma_i^- U(t)$. On the other hand, the Heisenberg equation for an operator $a(t)$ acting on the atomic subspace is 
\begin{eqnarray}\nonumber \frac{d}{dt}a=-&&i[a,H_a] +\sqrt{\frac\gamma{2\pi}}\int_{\bar\om-\theta}^{\bar\om+\theta}d\om\,b_\om^\dagger(t)\\ &&\big[a,  e^{-i(\bar\omega -\om)t} \sigma_1^-(t)+e^{-i(\bar\om-\om)(t-\tau)}e^{i\Delta\phi} \sigma_2^-(t)\big]\nonumber
\\-\nonumber &&\sqrt{\frac\gamma{2\pi}}\int_{\bar\om-\theta}^{\bar\om+\theta}d\om\,\Big[a,  e^{i(\bar\omega -\om)t} \sigma_1^+(t)\\ &&+e^{i(\bar\om-\om)(t-\tau)}e^{-i\Delta\phi} \sigma_2^+(t)\Big]b_\om(t).\label{heisenbergsys} \end{eqnarray}

Note that, in order to simplify the notation, we remove the time dependence of $a$ whenever it should be understood as $a(t)$. Formally integrating Eq.\,\eqref{heisenbergbom} and inserting the corresponding expression into Eq.\,\eqref{heisenbergsys}, we obtain 
\begin{equation}\begin{aligned} \frac{da}{dt}=-&i[a,H_a]+ \sqrt{\gamma}\big(\big[a,\xi^\dagger(t)\sigma_1^-(t)-\text{H.C.}\big]\big) \\+&\sqrt{\gamma}\big([a,\xi^\dagger(t-\tau) e^{i\Delta\phi}\sigma_2^-(t)-\text{H.C.}\big]\big)
\\+& \frac{\gamma}{2\pi}\int_{\bar\om-\theta}^{\bar\om+\theta}d\om\int_0^tdt'\,e^{-i(\bar\om-\om)(t-t')}\sigma_1^+(t') \\&\big[a,\sigma_1^-(t)  +e^{i(\bar\om-\om)\tau}e^{i\Delta\phi}\sigma_2^-(t)\big]
\\-& \frac{\gamma}{2\pi}\int_{\bar\om-\theta}^{\bar\om+\theta}d\om\int_0^tdt'\,\big[a,\sigma_1^+(t)  \\&+e^{-i(\bar\om-\om)\tau}e^{-i\Delta\phi}\sigma_2^+(t)\big]e^{i(\bar\om-\om)(t-t')}\sigma_1^-(t')
\\+& \frac{\gamma}{2\pi}\int_{\bar\om-\theta}^{\bar\om+\theta}d\om\int_0^tdt'\,e^{-i(\bar\om-\om)(t+\tau-t')}e^{-i\Delta\phi}\sigma_2^+(t')\\&\big[a,\sigma_1^-(t) +e^{i(\bar\om-\om) \tau}e^{i\Delta\phi}\sigma_2^-(t)\big]
\\-& \frac{\gamma}{2\pi}\int_{\bar\om-\theta}^{\bar\om+\theta}d\om\int_0^tdt'\,e^{i(\bar\om-\om)(t+\tau-t')}e^{i\Delta\phi}\big[a,\sigma_1^+(t) \\&+e^{-i(\bar\om-\om) \tau}e^{-i\Delta\phi}\sigma_2^+(t)\big]\sigma_2^-(t'),
\end{aligned}\end{equation}
where $\xi(t)=\frac{1}{\sqrt{2\pi}}\int_{\bar\om-\theta}^{\bar\om+\theta}d\om\,b_\om(0)e^{i(\bar\omega -\om)t}$ are the quantum noise operators. We now perform a Born-Markov treatment where the integration over $\om$ of the phase factors generates Dirac delta functions of $t'$, which allows to evaluate the integral in $t'$. This approximation is valid if $\sigma_i^-(t')\approx\sigma_i^-(t)$ for $t'\in[t-1/\theta,t+1/\theta]$, which requires $\gamma,\Omega,|\delta_i|\ll\theta$. This gives rise to terms such as \be\begin{aligned} &\int_{\bar\om-\theta}^{\bar\omega+\theta}d\om\int_0^tdt'\,e^{-i(\bar\om-\om)(t-t')}\sigma_1^+(t')[a,\sigma_1^-(t)] \\&\approx 2\pi\int_0^tdt' \delta(t-t')\sigma_1^+(t')[a,\sigma_1^-(t)]\\&=\pi\sigma_1^+(t)[a,\sigma_1^-(t)],\end{aligned}\ee
and \be\begin{aligned} &\int_{\bar\om-\theta}^{\bar\omega+\theta}d\om\int_0^tdt'\,e^{-i(\bar\om-\om)(t-t'-\tau)}e^{i\Delta\phi}\sigma_1^+(t')[a,\sigma_2^-(t)] \\&\approx 2\pi \int_0^tdt' \delta(t-t'-\tau)e^{i\Delta\phi}\sigma_1^+(t')[a,\sigma_2^-(t)]\\&=2\pi e^{i\Delta\phi}\sigma_1^+(t-\tau)[a,\sigma_2^-(t)].\end{aligned}\ee

For now we are interested in the Markovian limit where $\tau$ is set to $0^+$. This requires that the delay is much shorter than the typical evolution time of the system, i.e. $\gamma, \Omega, |\delta_i| \ll 1/\tau$, which we will assume from now on. We then obtain the quantum Langevin equation for the atomic operators
\begin{eqnarray} \frac{d}{dt}a=&-&i[a,H_a] \\&+&\nonumber \sqrt{\gamma} [a,\xi^\dagger(t)\sigma_1^-(t)-\text{H.C.}]\\&+&\nonumber\sqrt{\gamma}[a, \xi^\dagger(t)e^{i\Delta\phi}\sigma_2^-(t)-\text{H.C.}]\\
&+&\nonumber\frac{\gamma}2{\big(\sigma_1^+(t)[a,\sigma_1^-(t)] - [a,\sigma_1^+(t)]\sigma_1^-(t) \big)}
\\&+&\nonumber\frac{\gamma}2{\big(\sigma_2^+(t)[a,\sigma_2^-(t)] - [a,\sigma_2^+(t)]\sigma_2^-(t) \big)}
\\ &+&\nonumber \gamma{\big(e^{i\Delta\phi}\sigma_1^+(t)[a,\sigma_2^-(t)] - [a,\sigma_2^+(t)]e^{-i\Delta\phi}\sigma_1^-(t)\big)}.\end{eqnarray}

\begin{figure}
\includegraphics[width=9cm]{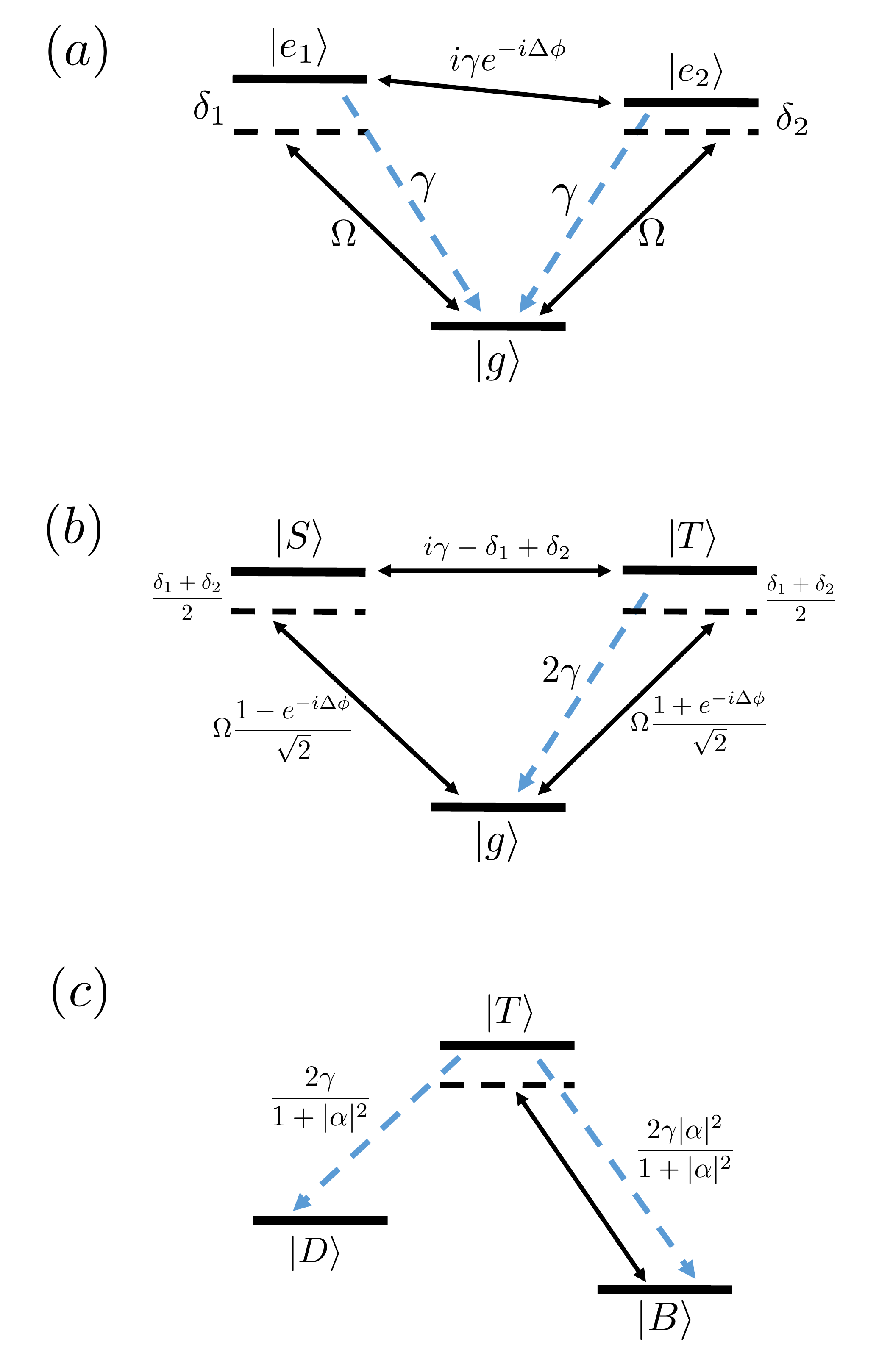}
\caption{\label{VsystemFig}(Color online)  Representation of the atom's dynamics in different basis. The coherent component is represented in solid black arrows and the incoherent component in dashed blue ones. (a) $\{\ket{g},\ket{e_1},\ket{e_2}\}$ basis. The dynamics is given by Eq.\,\eqref{eq:masterequationnonfinal}. The transitions are driven with a Rabi frequency $\Omega$ and undergo a dipole-dipole interaction. (b) $\{\ket{g},\ket S,\ket T\}$ basis. The dynamics is given by Eq.\,\eqref{eq:vsystemmasterequation}. The state $\ket T$ deexcites at a rate $2\gamma$ and is coherently coupled to the other two states. When $\Delta\phi=0$, the state $\ket{S}$ is decoupled from the ground state. (c) $\{\ket{D},\ket{B},\ket{T}\}$ basis. The dynamics coherently couples the bright component $\ket{B}$ to the decaying state $\ket{T}$, which incoherently jumps to the other states via photon emission. Eventually the system is pumped into the dark state $\ket{D}$.}
\end{figure}

Let us write down the expectation value of this equation for an initial state $\ket{\Psi_0}$ where the waveguide is in the vacuum state. In that case, $\xi(t) \ket{\Psi_0}=\bra{\Psi_0}\xi^\dagger(t) =0$, and we get
 \begin{eqnarray}\label{eq:averageeq}
\frac{d\average{a}}{dt}=&-&i\average{[a,H_a]}\\&+&\nonumber\frac{\gamma}2\average{\sigma_1^+(t)[a,\sigma_1^-(t)] - [a,\sigma_1^+(t)]\sigma_1^-(t) }
\\&+&\nonumber\frac{\gamma}2\average{\sigma_2^+(t)[a,\sigma_2^-(t)] - [a,\sigma_2^+(t)]\sigma_2^-(t) }
\\ &+&\nonumber \gamma\average{e^{i\Delta\phi}\sigma_1^+(t)[a,\sigma_2^-(t)] - [a,\sigma_2^+(t)]e^{-i\Delta\phi}\sigma_1^-(t)}
\end{eqnarray}
We now move to the Schr\"odinger picture and express the average terms in Eq.\,\eqref{eq:averageeq} as \be \average{a}=\bra{\Psi_0}a\ket{\Psi_0}=\text{Tr}_a\big(\text{Tr}_w(a(0)\ket{\Psi(t)}\bra{\Psi(t)})\big),\ee where $\text{Tr}_w$ denotes the trace over the waveguide modes, and $\text{Tr}_a$ the trace over the atomic states. The atomic density matrix $\rho$ is obtained from the full density matrix $\ket{\Psi(t)}\bra{\Psi(t)}$ by tracing over the waveguide modes: $\rho(t)=\text{Tr}_w(\ket{\Psi(t)}\bra{\Psi(t)})$. Notice that every operator appearing in Eq.\,\eqref{eq:averageeq} acts only on the atomic Hilbert space, and can thus be taken out of the waveguide trace $\text{Tr}_w$. The average terms thus read $\average{a}=\text{Tr}_a(a(0)\rho(t))$, and by using the cyclic property of the trace, one finally obtains the master equation. 
\begin{equation}\label{eq:masterequationnonfinal} \frac{d\rho}{dt}=-i[H_a+H_\text{dd},\rho]+\gamma\mathcal D[\sigma_1^-+\sigma_2^-]\rho, \end{equation}
where the effective dipole-dipole interaction term reads \be H_\text{dd}\equiv i\frac{\gamma}2\big(e^{i\Delta\phi}\sigma_1^+\sigma_2^--e^{-i\Delta\phi}\sigma_2^+\sigma_1^-\big).\ee
We will denote the effective Hamiltonian as $H_\text{eff}=H_a+H_\text{dd}$. In Fig.\,\ref{VsystemFig}(a) we show the level scheme of the system along with the different terms of the master equation.

It is convenient to introduce the following states \be\begin{aligned}\ket{S}&=\frac{1}{\sqrt{2}}\big(\ket{e_1}-e^{-i\Delta\phi}\ket{e_2}\big), \\\ket{T}&=\frac{1}{\sqrt{2}}\big(\ket{e_1}+e^{-i\Delta\phi}\ket{e_2}\big),\label{eq:defST}\end{aligned}\ee and the corresponding operators $\sigma_S^- =\ket{g}\bra{S}$ and $\sigma_T^-=\ket{g}\bra{T}$. The master equation can then be expressed as  \begin{equation}\label{eq:vsystemmasterequation} \frac{d\rho}{dt}=-i[H_\text{eff},\rho]+2\gamma\mathcal D[\sigma_T^-]\rho, \end{equation}
where the Hamiltonian now reads \be\begin{aligned} H_\text{eff} = &-\frac{\delta_1+\delta_2}2 \big(\ket{S}\bra{S}+\ket{T}\bra{T}\big) \\&+ \Big(\frac{i\gamma-\delta_1+\delta_2}2\sigma_S^+\sigma_T^-+\text{H.C.}\Big)
\\ &-\frac{\Omega}{2}\Big(\frac{1-e^{-i\Delta\phi}}{\sqrt{2}} \sigma_S^- + \frac{1+e^{-i\Delta\phi}}{\sqrt{2}} \sigma_T^- + \text{H.C.}\Big).\end{aligned}\ee
We see from Eq.\,\eqref{eq:vsystemmasterequation} that the state $\ket{T}$ is \emph{superradiant} as it decays with a rate $2\gamma$, while the state $\ket{S}$ is \emph{subradiant} as it does not spontaneously decay. Although the latter state does not generate radiation, it is not a dark state since it is unstable, due to the dipole-dipole term. We illustrate the dynamics of the atom in Fig.\,\ref{VsystemFig}(b).

Notice that the master equation of Eq\,\eqref{eq:vsystemmasterequation} is very similar to the equation for the cascaded two-level atoms of Sec.\,\ref{sec:dimers} (Eq.\,\eqref{eq:dimermasterequation}), where the state $\ket{g_1g_2}$ is now replaced by $\ket{g}$, the state $\ket{e_1g_2}$ by $\ket{e_1}$ and the state $\ket{g_1e_2}$ by $\ket{e_2}$. The double-excited state $\ket{e_1e_2}$ however does not have an equivalent in our feedback setup.

\subsection{Formation of dark states}
\label{subsec:darkstates}
Due to the dissipation of the open system via photon emission, the system relaxes towards a steady-state which is in general mixed. However, depending on the settings of the driving laser, this steady-state can be a pure state $\ket D$. The two conditions for the existence of such a state are \cite{PhysRevA.78.042307}

\begin{enumerate}
\item $\big[H_\text{eff},\ket{D}\bra{D}\big]=0$,
\item $\sigma_T^-\ket{D}=0$.
\end{enumerate}

The first condition implies that the state is stationary, i.e. an eigenstate of $H_\text{eff}$. The second condition requires that no incoherent stochastic jump occurs, which is a requirement for the state to be pure. Since the jump operator is $\sigma_T^-$, such a state belongs to the manifold spanned by $\ket g$ and $\ket S$. We can then write \be \ket D=\frac{1}{\sqrt{1+|\alpha|^2}}\big(\ket g + \alpha \ket S\big) \ee for some $\alpha$, in analogy with Eq.\,\eqref{eq:dimerD}. Requiring this state to be an eigenstate of $H_\text{eff}$ provides two constraints. Denoting the projector on the $\{\ket{g},\ket{S}\}$ subspace $P=\ket{g}\bra{g}+\ket{S}\bra{S}$, the first one is $(1-P)H_\text{eff}\ket{D}=0$ which means that the coupling between $\ket{D}$ and $\ket{T}$ must vanish, and yields
\begin{equation}\label{alpha1rstcond}
{\alpha=-\frac{\Omega}{\sqrt{2}}\frac{1 + e^{i\Delta\phi}}{i\gamma+\delta_1-\delta_2}.}\end{equation}
The second one reads $PH_\text{eff}\ket{D} \propto \ket{D}$ which means that $\ket{D}$ is an eigenstate of the effective Hamiltonian restricted to the $P$ subspace, and yields
\begin{equation}\label{alpha2ndcond}{\begin{aligned}\alpha^2\Omega(&e^{-i\Delta\phi}-1)/\sqrt{2}+\alpha(\delta_1 + \delta_2) \\ +&\Omega(1-e^{i\Delta\phi})/\sqrt{2} =0.
\end{aligned}}\end{equation}

The existence of a solution for $\alpha$ satisfying these two constraints strongly depends on the phase difference $\Delta\phi$. In what follows, we will consider two different regimes, namely when the \emph{commensurability condition} $\Delta\phi=0$ is satisfied, and when it is not. Note that if one drives the atom through the waveguide, $\phi'=\pi-\bar\omega \tau$ and the commensurability condition is automatically satisfied. In order to tune $\Delta\phi$ to different values, the driving must thus be done using fields outside of the waveguide. We now show that dark states can arise in both regimes and that they display different properties.

\subsubsection{$\Delta\phi=0$}

In Sec.\,\ref{sec:dimers}, we have assumed this condition satisfied. Indeed, one can show \cite{PhysRevA.91.042116} that for the quantum dimers, any deviation from this phase induces a coherent coupling between the $\ket{S_{12}}$ state from Eq.\,\eqref{eq:dimerD} and the double-excited state $\ket{e_1e_2}$. Since the latter state is not destroyed by the Lindblad jump operator, the commensurability is a necessary condition for a dark steady-state. In our case, if we assume this condition satisfied, Eq.\,\eqref{alpha2ndcond} simply becomes $\delta_1+\delta_2=0$, signifying that the state $\ket{S}$ is not detuned from $\ket{g}$. The $\ket{S}$ fraction is then given by \be \label{alphacommens} \alpha=-\sqrt{2}\frac{\Omega}{i\gamma+2\delta_1}\ee which increases \emph{linearly} with the Rabi frequency. For {any} $\Omega$, there exists a dark state as the unique steady-state of the dynamics. Note that this is the same expression as Eq.\,\eqref{eq:alphadimer}, which shows the similarities between the physics of our system and the quantum dimer.

We define the \emph{bright} state $\ket{B}=\frac{1}{\sqrt{1+|\alpha|^2}}(\alpha^*\ket{g}-\ket{S})$, which is bright in the sense that, contrary to the dark state, it is coupled to $\ket{T}$ which eventually decays by emitting photons. The three states $\{\ket{T},\ket{D},\ket{B}\}$ form an orthonormal basis on the atomic Hilbert space, and the dynamics is represented in Fig.\,\ref{VsystemFig}(c), where it is clear that $\ket{D}$ is the steady-state. On this basis, the decay from $\ket{T}$ to $\ket{g}$ generates an effective decay to $\ket{D}$ with a rate $\gamma_\text{eff}=2\gamma/(1+|\alpha|^2)$ and to $\ket{B}$ with a rate $|\alpha|^2\gamma_\text{eff}$. The time necessary to reach the steady-state is roughly given by $2\pi/\gamma_\text{eff}$, which grows quadratically with $\Omega$. The Hamiltonian from Eq.\,\eqref{eq:vsystemmasterequation} now takes the simple form \be H_\text{eff}=\Big(i\frac{\gamma}{2} + \delta_1\Big)\sqrt{1 + \frac{2 \Omega^2}{|i \gamma + 2 \delta_1|^2}} \ket{T}\bra{B}.\ee

\subsubsection{$\Delta\phi\neq0$}

The fact that the state $\ket{e_1e_2}$ from the quantum dimer setup has no equivalent in our system allows to construct dark states even when the commensurability is not satisfied, i.e. the feedback photons are out of phase with the driving. Inserting Eq.\,\eqref{alpha1rstcond} into Eq.\,\eqref{alpha2ndcond}, a relation between the variables of the system can be obtained. We require $\Omega$ to be real, and find that the detunings must satisfy $\delta_1 - \delta_2 = 0$, which is the exact opposite condition as for the $\Delta\phi=0$ case. From now on we will assume this condition to be satisfied, and we will denote $\delta = \delta_1=\delta_2$. The requirement for obtaining a dark state then becomes \begin{equation} \label{eq:OmegaResultsymmetric} \Omega/\gamma=\sqrt{\frac{1}{1+\cos(\Delta\phi)}- \frac{2 \delta/\gamma}{\sin(\Delta\phi)}}. \end{equation}
This dark state can be interpreted as follows (see Fig.\,\ref{VsystemFig}(b)). In contrast to the case $\Delta\phi=0$, the two states $\ket{g}$ and $\ket{S}$ are coupled by the coherent part of the dynamics. Since a pure steady-state has to be an eigenstate of this coherent evolution, it restricts the possibility to the two dressed states $\ket{\pm} = (\ket{g}+\alpha_\pm\ket{S})/\sqrt{1+|\alpha_\pm|^2}$, where \be \label{eq:alphapm}\alpha_\pm=-\frac{\Omega\frac{1-e^{i\Delta\phi}}{\sqrt{2}}}{\delta\pm\text{sgn}(\Delta\phi)\sqrt{\delta^2+|\Omega\frac{1-e^{i\Delta\phi}}{\sqrt{2}}|^2}}.\ee Eq.\,\eqref{eq:OmegaResultsymmetric} then states the condition under which the couplings to the state $\ket{T}$ from the components of the state $\ket{g}$ and the state $\ket{S}$ in the state $\ket{+}$ interfere destructively, rendering $\ket{+}$ a dark eigenstate of the coherent dynamics with eigenvalue \begin{equation}\label{eqE+} E_+ = -\delta + \frac{\gamma}2\frac{1-\cos(\Delta\phi)}{\sin(\Delta\phi)}. \end{equation} In particular, for $\delta=0$ we find $|\alpha|=1$, which is now \emph{independent} of the Rabi frequency. Moreover, from Eq.\,\eqref{eq:OmegaResultsymmetric} we note that this dark state appears only above a critical Rabi frequency $\Omega_c = \gamma/\sqrt{2}$.

The bright state $\ket{B}$ is identified as the other dressed state $\ket{-}$ with the energy \be E_- = \label{eqE-}- \frac{\gamma}2\frac{1-\cos(\Delta\phi)}{\sin(\Delta\phi)}.\ee  On the $\{\ket{T},\ket{D},\ket{B}\}$ basis, the coherent part of the dynamics of the system, represented by the black arrow in Fig.\,\ref{VsystemFig}(c), is now governed by the Hamiltonian \be\begin{aligned} H_\text{eff} =& -\delta\ket{T}\bra{T}+E_+\ket{D}\bra{D}+E_-\ket{B}\bra{B}
\\ &+ \big(J_{TB}\ket{T}\bra{B}+\text{H.C.}\big),\end{aligned}\ee
where the coupling can be expressed as  \be J_{TB}= i\frac\gamma2\sqrt{2+\delta/E_-}. \ee
In particular, if $\Delta\phi\to\pi$ then $E_\pm\to\pm\infty$ whereas $J_{TB}$ remains finite. In this limit, we see that all three states effectively decouple, hence the time necessary to reach the steady-state $\ket{D}$ diverges. For other values of $\Delta \phi$ however, this time remains finite.

\begin{figure}
\includegraphics[width=9.cm]{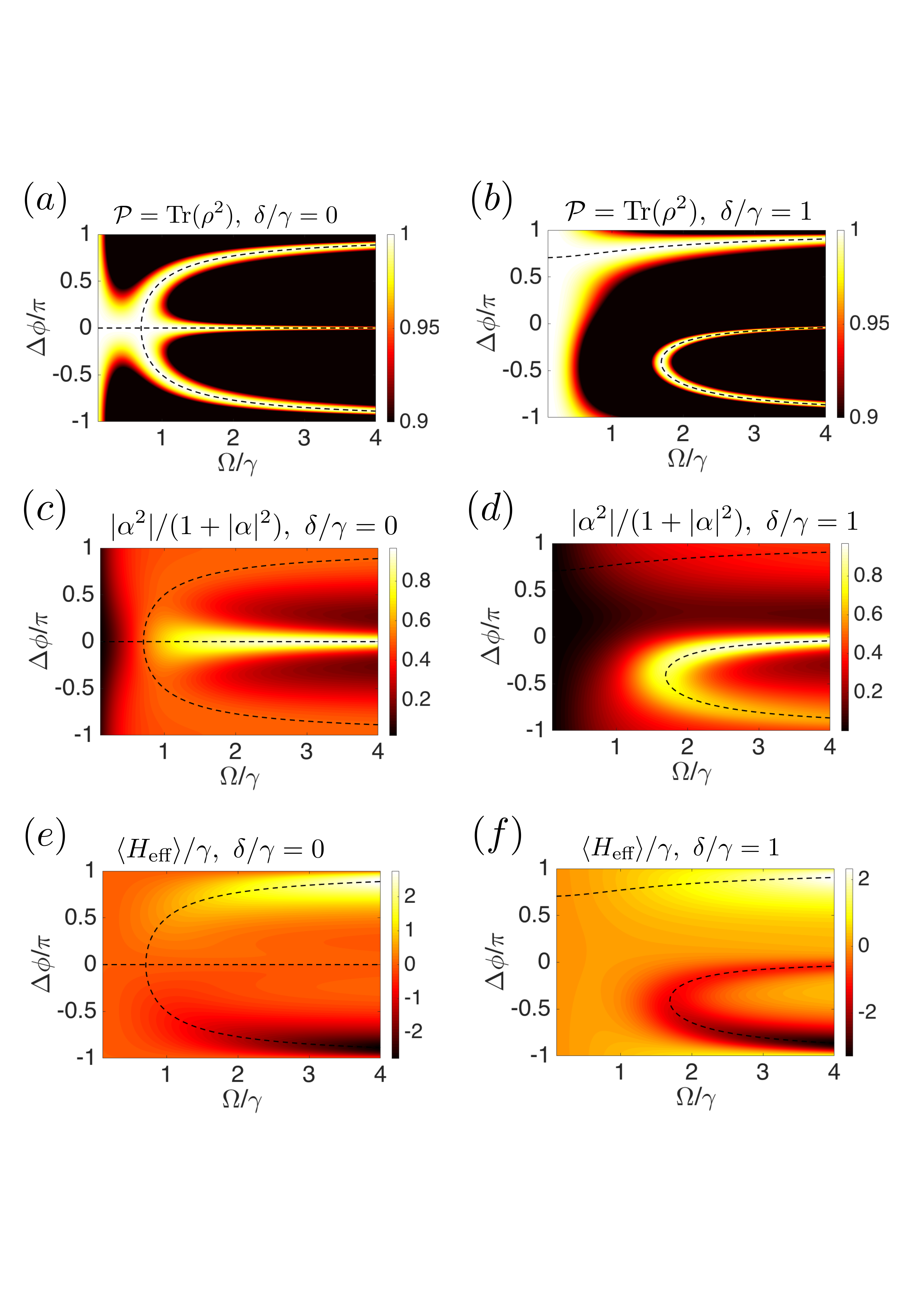}
\caption{\label{fig:markov_purity}(Color online)  Properties of the steady-state, as a function of the Rabi frequency $\Omega$ and the phase difference $\Delta\phi$. The master equation \eqref{eq:vsystemmasterequation} was solved exactly by vectorising $\rho$. In dashed black lines, we plot the curves along which we predicted dark states. In the first column, $\delta=0$, whereas $\delta=\gamma$ in the second one. In the first row we plot the purity, which is exactly 1 along the black lines. In the second row we plot the excitation probability, which is the occupancy of $\ket{S}$. In the third row we plot the expectation value of $H_\text{eff}$ in units of $\gamma$.}
\end{figure}

We now look at the whole parameter range for $\Omega$ and $\Delta\phi$. In Fig.\,\ref{fig:markov_purity}, we plot some of the properties of the steady-state as a function of $\Omega/\gamma$ and $\Delta\phi$. When $\delta=0$, the horizontal dashed black line corresponds to the case $\Delta\phi=0$. The other dashed curve is a plot of Eq.\,\eqref{eq:OmegaResultsymmetric}. In Fig.\,\ref{fig:markov_purity}(a), we see that for any given phase $\Delta\phi\neq(0,\pi)$ there exists a unique Rabi frequency $\Omega$ for which the steady-state is dark. By adiabatically increasing the laser intensity from zero, one would thus observe a dip in the intensity of the photon emission in the waveguide, which is the signature of the dark state. Conversely, for $\Omega\leq\Omega_c=\gamma/\sqrt{2}$ the only possible phase is $\Delta\phi=0$ whereas for $\Omega>\Omega_c$ three different values lead to a dark state. Fig.\,\ref{fig:markov_purity}(c) shows that, as predicted, the occupation of $\ket{S}$ along the $\Delta\phi=0$ line rapidly converges to 1 when $\Omega/\gamma$ increases, whereas along the other curve it remains constant at 1/2. In Fig.\,\ref{fig:markov_purity}(e), we see that the energy of the dark state has the same sign as $\Delta\phi$. When the detuning becomes non-zero (Figs.\,\ref{fig:markov_purity}(b,d,f)), both black curves merge into two bands, which are now separated by a phase gap. 

\subsection{Coupling to a cavity in the bad-cavity regime}
 \label{sec:couplcavi}

In the previous sections we have always assumed that the coupling of the atom to non-guided modes $\gamma'$ was negligible compared to the coupling to the guided ones $\gamma$. This is currently not the case when working with real atoms, where $\beta=\gamma/(\gamma+\gamma')\lesssim0.5$ \cite{hood2016} in setups where the atom-waveguide coupling is not chiral, and $\beta\approx 0.025$ \cite{PhysRevX.5.041036} in setups featuring chiral couplings. A way to increase the effective atom-waveguide cooperativity is depicted in Fig.\,\ref{fig:vsyswithwaveguide}(c), where the atom couples with a rate $g$ to a cavity which preserves chiral coupling \cite{PhysRevLett.110.213604}. The cavity is resonant with the laser and is coupled to the guided modes with a rate $\kappa$.

 In a frame rotating at the laser frequency, the master equation for the system consisting of atom and cavity is given by
\begin{equation}\label{eq:masteqrhoca} \dfrac{d\rho}{dt} = -i[H_a,\rho] + \mathcal{L}'\rho+\mathcal L_\text{int}\rho+\mathcal L_\text{cav}\rho ,\end{equation}
where the loss to non-guided modes from the atom is given by \begin{equation} \mathcal L'\rho=\gamma' \big( \mathcal D[\sigma_1^-]\rho + \mathcal D[\sigma_2^-]\rho\big), \end{equation}
 the coupling between the cavity and the atom is described by \begin{equation} \label{eq:Lidef} \mathcal L_\text{int} \rho = -ig[\sigma^+_T a_T +\sigma_S^+ a_S + \text{H.C.},\rho]\end{equation}
and the free evolution of the cavity modes is given by \begin{eqnarray} \label{eq:defLB}  \mathcal L_\text{cav}\rho = &&\frac{\kappa}{2}[a^\dagger_S a_T - a^\dagger_T a_S,\rho]+2\kappa\mathcal D[a_T]\rho \nonumber \\ &&+ \kappa'\big(\mathcal D[a_1]\rho+\mathcal D[a_2]\rho\big).\end{eqnarray} Here, $\kappa'$ is the intra-cavity loss rate due to absorption and coupling to non-guided modes, $a_1$ and $a_2$ are the annihilation operator for the cavity modes coupled respectively to the $\sigma_1$ and $\sigma_2$ transitions, and we have defined the $T$ and $S$ cavity modes $a_{T/S} = (a_1 \pm e^{i\Delta\phi}a_2)/\sqrt{2}$ in analogy with the $\sigma_T^-$ and $\sigma_S^-$ atomic operators. In order to write down the master equation, we assumed that the time-delay $\tau$ is much smaller than the relevant timescale of the system, namely here $1/\kappa$. 

 As detailed in Appendix \ref{appendixcavity}, in the bad-cavity regime $g\ll\kappa$ the cavity can be adiabatically eliminated \cite{Cirac:1992fl}, and the density matrix for the atom $\rho_a$, obtained from $\rho$ by tracing over the cavity modes, is governed by the following master equation
 \be  \label{eqcavityadiab}\begin{aligned}\frac{d\rho_a}{dt}=&-i[H_a,\rho_a]+2\gamma\mathcal{D}[\sigma_T^-]\rho_a \\&+\frac{\gamma}2[\sigma_S^+\sigma_T^--\sigma_T^+\sigma_S^-,\rho_a]\\&+ (\gamma'+\gamma\kappa'/\kappa)\big(\mathcal D[\sigma_1^-]\rho_a+\mathcal D[\sigma_2^-]\rho_a\big),\end{aligned}\ee
where $\gamma=(2g)^2\kappa/(\kappa+\kappa')^2$ is identified as the effective atom-waveguide coupling and $\Delta\phi$ has been redefined with an additional $\pi$-shift, and hence now reads $\Delta\phi=-\bar\omega \tau-\phi'$. This shift is reminiscent of the cavity and the fact that a resonant photon entering it will leave with a $\pi$-shift \cite{Volz:2014aa}. The dynamics of Eq.\,\eqref{eqcavityadiab} is equivalent to the case without a cavity (Eq.\,\eqref{eq:vsystemmasterequation}), with an additional coupling to non-guided modes whose rate is identified as $\gamma'_\text{tot}=\gamma'+\gamma\kappa'/\kappa$, so that the ratio between the respective couplings to the non-guided and to the guided modes is \be \frac{\gamma'_\text{tot}}{\gamma}=\frac{\gamma'}{\gamma}+\frac{\kappa'}{\kappa}. \ee
 
 In order to have a low ratio, one needs $\kappa'\ll\kappa$ (in which case $\gamma\approx (2g)^2/\kappa$) and $\gamma'\ll\gamma$.

 \section{Beyond the Markovian approximation}\label{section:nonmarkov}
 
 In this section we study the dynamics of our setup beyond the Markovian regime, when the retardation effects in the coherent quantum feedback become important. We are thus interested in the regime where the time-delay $\tau$ is non-negligible compared to the relevant timescales, namely $1/\Omega$, $1/\gamma$, $1/|\delta_i|$. Due to the constant driving, a non-negligible photon number is now present between the atom and the mirror, which results in entanglement between these photons and the atom, and in a retarded dipole-dipole interaction between the two transitions of the atom. The previous master equation treatment is no longer valid \cite{PhysRevLett.116.093601,PhysRevB.76.045317} and one needs to employ numerical methods in order to track the entangled states of the electromagnetic field and of the atom. Recently, techniques have been developed to study the dynamics of photonic circuits consisting of quantum optical systems coupled via waveguides, where the time-delay in the interaction can be significant \cite{PhysRevLett.116.093601}. The state of the system comprising the atom and the photonic field is approximated using matrix-product states (MPS) methods \cite{RevModPhys.77.259}, which are well suited to account for the entanglement in our system. We provide a brief description of the method in Appendix \ref{appendixb}. In this section we first neglect the effect of the non-guided modes and assume a perfect chiral coupling.

 \begin{figure}
\includegraphics[width=9cm]{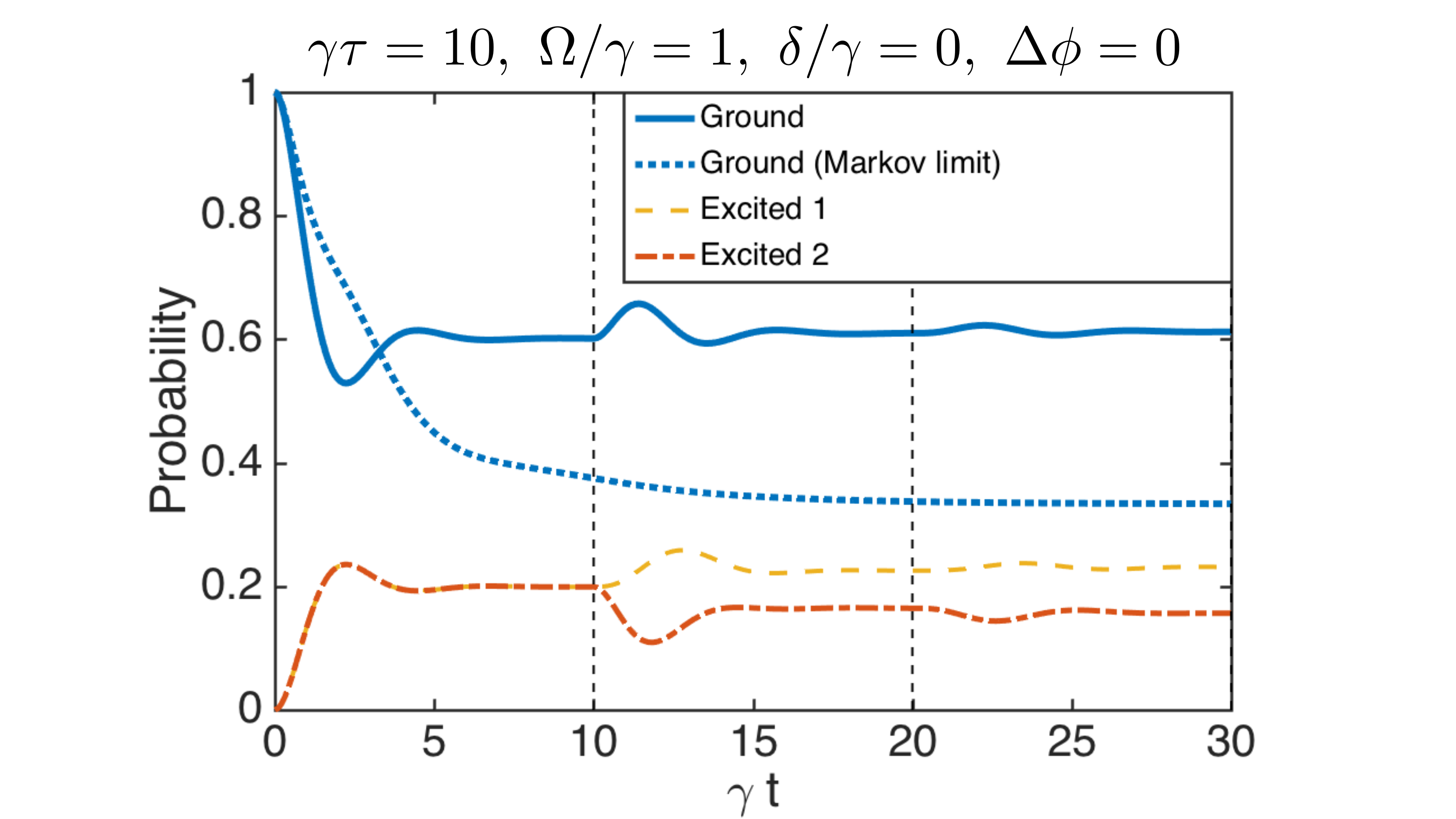}
\caption{\label{fig:longdelay} (Color online) Atomic occupation probabilities as a function of time for a long delay $\tau = 10/\gamma$, with $\Omega=\gamma, \delta=0$ and $\Delta\phi=0$. For reference we plot the ground state occupation in the Markovian limit.}
\end{figure} 
 \begin{figure}
\includegraphics[width=8.8cm]{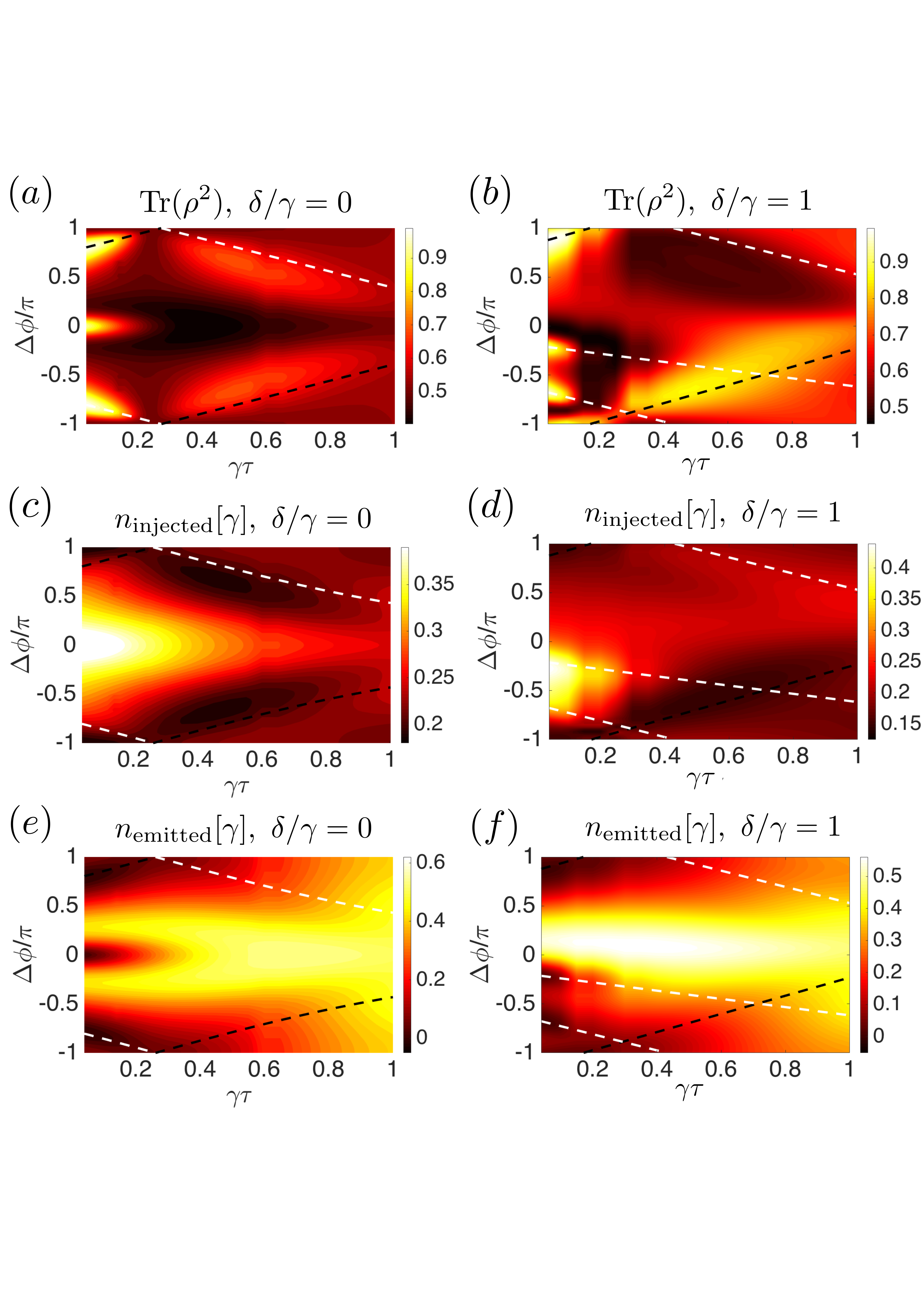}
\caption{\label{fig:nonmarkov_purity}(Color online)  Properties of the steady-state as a function of the phase difference $\Delta\phi$ and the propagation time $\gamma \tau$, for $\Omega=2\gamma$ and $\delta=0$ (first column) or $\delta=\gamma$ (second column). In the first row we plot the purity of the reduced density matrix of the atom. In the second row we plot the number of photons injected inside the feedback loop per unit time $1/\gamma$ via the $\sigma_1$ transition. In the third row we plot the number of photons emitted in the waveguide, outside of the feedback loop, per unit time $1/\gamma$. In dashed lines we plot $\phi_D + (E_+ - E_-)\tau$, where $\phi_D$ is one of the values of $\Delta\phi$ for which the steady-state is dark in the Markovian limit, and $E_+$ and $E_-$ are the energies of the corresponding dark and bright states, given by Eq.\,\eqref{eqE+} and Eq.\,\eqref{eqE-}, in black for $\Delta\phi>0$ and in white for $\Delta\phi<0$.   }
\end{figure}
 
 In Fig.\,\ref{fig:longdelay} we show the evolution of the populations of the atomic reduced density matrix as a function of time. Up to time $t=\tau=10/\gamma$, the system evolves freely, as if the mirror was not present. The solution is thus given by a Rabi oscillation between the ground state $\ket{g}$ and the state $(\ket{e_1}+\ket{e_2})/\sqrt{2}$ with a Rabi frequency $\sqrt{2}\,\Omega$, and a dissipation induced by photon emission into the waveguide. At time $\tau$ the system starts interacting with the feedback, and for $\Delta\phi=0$, the feedback photons previously emitted by the $\sigma_1$ transition are perfectly in phase with the photons emitted by the $\sigma_2$ transition. This generates a constructive photon interference which amplifies the emission via a superradiance process. This is demonstrated by the sudden dip in the excitation of $\ket{e_2}$ and the bump for the ground state. Consequently, a fraction of this ground state bump will be transferred to the occupation of $\ket{e_1}$ by the laser driving which leads to the bump in the excitation of $\ket{e_1}$. A similar process can be distinguished for $t=2\tau$, after which the system becomes very close to the steady-state.
 
 In Fig.\,\ref{fig:nonmarkov_purity}, we show how some of the steady-state properties are affected by an increasing time-delay $\tau$. In Fig.\,\ref{fig:nonmarkov_purity}(a) and Fig.\,\ref{fig:nonmarkov_purity}(b), we see that the purity is locally maximal along the lines given by $\Delta\phi = \phi_D + (E_+ - E_-)\tau$, where $\phi_D$ is one of the values of $\Delta\phi$ for which the steady-state is dark in the Markovian limit, and $E_+$ and $E_-$ are the energies of the corresponding dark and bright states, which are given by Eq.\,\eqref{eqE+} and Eq.\,\eqref{eqE-}. For $\delta=0$, the phase of the solution $\phi_D=0$ is not shifted. The other two solutions are shifted symmetrically and thus cross at $\Delta\phi=\pm\pi$. The purity decreases locally around this point, which indicates that the steady-state is now mixed. Figs.\,\ref{fig:nonmarkov_purity}(c,e) show that the photon number in the waveguide is however still very low (albeit non-zero). For larger values of $\gamma \tau$, the photon emission increases along the dashed lines, therefore the purity vanishes. The fact that the photon emission in the dark state with $\phi_D=0$ increases much more rapidly with $\gamma\tau$ than for the other dark states is related to the fact that the component of the dark state which lies in the excited manifold, given by $\alpha$, is higher, as can be seen from Fig.\,\ref{fig:markov_purity}(c). This leads to a higher number of photons injected in the feedback loop, as represented in Fig.\,\ref{fig:nonmarkov_purity}(c), hence even a small delay generates an important number of feedback photons which are entangled with the system. Similar results are obtained with $\delta=\gamma$, in Figs.\,\ref{fig:nonmarkov_purity}(b,d,f). Along the black line, the dark state is much more robust to the increase of the delay than for the white lines, which is due to the fact that the value of $\alpha$ is lower (see Fig.\,\ref{fig:markov_purity}(d)).

\section{Experimental considerations}\label{sec:experiment}
To conclude this work we discuss some experimental considerations for the physical implementation of our system.  

 \subsection{Effect of non-guided modes}

\begin{figure}
\includegraphics[width=8.8cm]{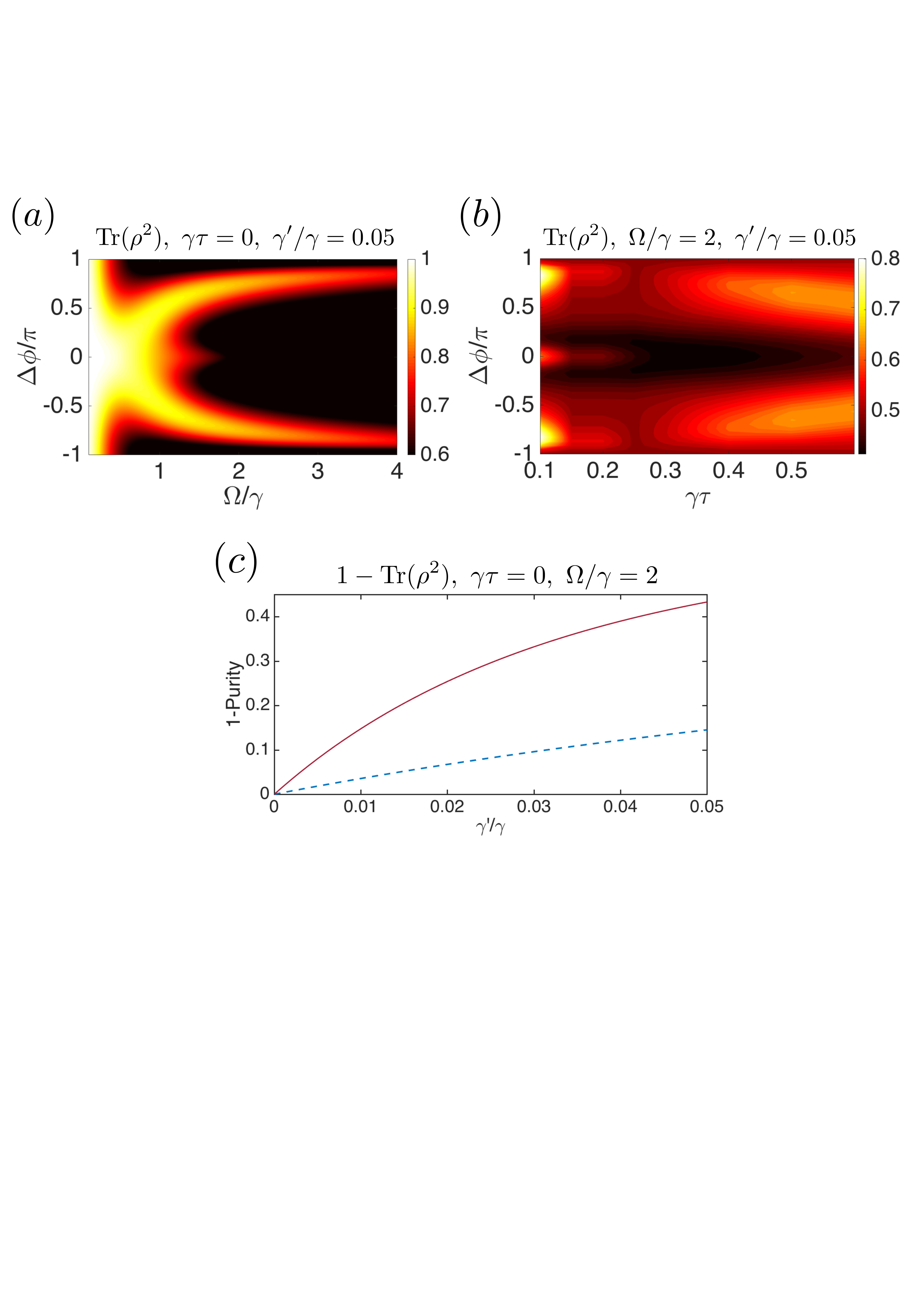}
\caption{\label{fig:decoherence}(Color online)  Effect of the coupling to non-guided modes. (a) Purity of the steady-state in the Markovian limit, as a function of $\Omega/\gamma$ and $\Delta\phi$, for $\gamma'/\gamma = 5\%$ and $\delta=0$. (b) Purity of the steady-state as a function of $\gamma\tau$ and $\Delta\phi$, for $\gamma'/\gamma=5\%$ and $\delta=0$. (c) Impurity of the steady-state in the Markovian limit for $\Omega/\gamma=2$, where $\Delta\phi$ is set to the dark state phases ($\Delta\phi =0$ in solid red, $\Delta\phi\neq0$ in dashed blue). }
\end{figure}

 We now consider the case where the atom can spontaneously deexcite by emitting a photon in the non-guided modes with a rate $\gamma'$. This coupling will effectively decrease the purity of the steady-state. In Fig.\,\ref{fig:decoherence} we investigate the robustness of the dark states. In the Markovian regime, this is done by adding a decoherence term $\gamma'(\mathcal D[\sigma_1^-]\rho+\mathcal D[\sigma_2^-]\rho)$ to the master equation (Eq.\,\eqref{eq:vsystemmasterequation}). Fig.\,\ref{fig:decoherence}(a) shows that in this regime, increasing $\Omega$ diminishes the purity. Two effects are in play here. For the dark state with $\Delta\phi=0$, we saw from Eq.\,\eqref{alphacommens} that $\alpha$ increases proportionally with the Rabi frequency. Since $\alpha$ is the excited fraction of the dark state, which can spontaneously emit in the non-guided modes, increasing $\Omega$ lowers the purity. For the dark states with $\Delta\phi\neq 0$, we saw that the time required to reach the steady-state diverges as $\Delta\phi$ approaches $\pm \pi$. As this time becomes large compared to the decoherence time $1/\gamma'$, the purity decreases. In Fig.\,\ref{fig:decoherence}(b) we show the effect in the non-Markovian regime. As the delay $\tau$ increases, the effect on the purity is not significantly increased, and the shifting of the phase with the delay is still very recognizable. In Fig.\,\ref{fig:decoherence}(c) we show the purity as a function of $\gamma'$ for $\Omega=2\gamma$. The dark state with $\Delta\phi\neq0$ (in dashed blue) is more robust than the one with $\Delta\phi=0$ (in solid red) which already for $\gamma'/\gamma =2\%$ has a purity of $\mathcal P \approx 0.75$.
 
 \subsection{Imperfection of the chiral coupling}
 
 Finally, we consider the case where the coupling between the atom and the waveguide modes is not perfectly chiral. Each transition of the atom $\sigma_1$ and $\sigma_2$ now couples to the guided modes propagating in both directions. Let us define the directionality of the coupling as $\eta=\gamma_\text{pref}/(\gamma_\text{pref}+\gamma_\text{imp})$, where $\gamma_\text{pref}$ is the decay rate in the preferred direction and $\gamma_\text{imp}=\gamma-\gamma_\text{pref}$ the decay rate in the other direction. We assume that both transitions have the same directionality, although their preferred directions are opposite. In the definition of the atom-waveguide interaction Hamiltonian (Eq.\,\eqref{eq:Hintdefdsa}), the creation of a photon propagating towards the mirror is now associated with the atomic operator $\sigma_L^- = \sqrt{\eta}\, \sigma_1^-+\sqrt{1-\eta}\,\sigma_2^-$ instead of $\sigma_1^-$. For a photon propagating outwards, the atomic operator is $\sigma_R^- = \sqrt{1-\eta}\,\sigma_1^-+\sqrt{\eta}\,\sigma_2^-$ instead of $\sigma_2^-$. 
 
In the Markovian regime, the master equation (Eq.\,\eqref{eq:vsystemmasterequation}) is now modified in the following way. First, the relative laser phase $\phi'$ can no longer be gauged away and is instead an independent parameter which we will set to zero. Second, the Lindblad jump operator is redefined as $\sigma_T^-=\sigma_L^-+e^{i\Delta\phi}\sigma_R^-$, and undergoes a dipole-dipole interaction with the redefined operator $\sigma_S^-=\sigma_L^--e^{i\Delta\phi}\sigma_R^-$.
 
\begin{figure}
\includegraphics[width=8.9cm]{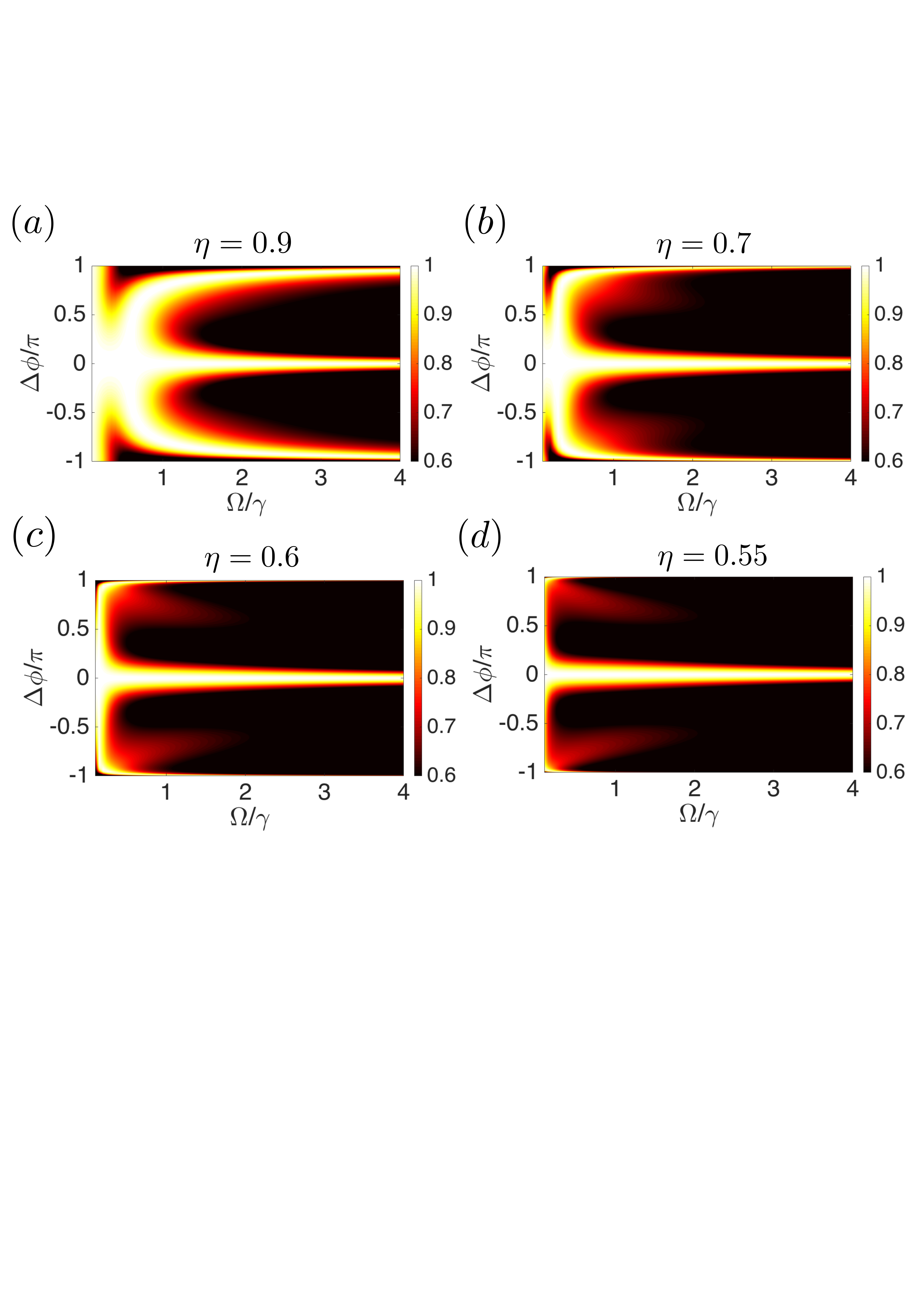}
\caption{\label{Direct}(Color online)  Effect of the directionality $\eta$ of the system on the purity of the steady-state for $\delta=0$ in the Markovian limit.  }
\end{figure}

 In Fig.\,\ref{Direct} we study the effect of this imperfect directionality. For the dark state with $\Delta\phi=0$, the directionality does not noticeably alter the purity. This situation is in fact analogous to the dimerization of two-level atoms coupled to a bidirectional waveguide, where now the atomic pair couples to the left-moving guided modes with a rate $\gamma_L$ and to the right-moving ones with a rate $\gamma_R$. This system has been investigated in Ref.\,\cite{PhysRevA.91.042116}, where the authors have shown that as long as $\gamma_L\neq\gamma_R$, the atoms dimerize to form a unique pure steady-state. In our system, $\gamma_L$ is identified as $\gamma_\text{pref}$ and $\gamma_R$ as $\gamma_\text{imp}$. 
 
 For $\Delta\phi\neq0$, on the contrary, the dark state vanishes as $\eta\to 0.5$. However, we see that the purity of the steady-state is not drastically altered for small variations of $\eta$. For example, for $\eta=0.9$, the results are still very similar to the predictions of the scenario with a perfect directionality $\eta=1$.

 \section{Conclusion}
 
 In this paper we have studied the dissipative dynamics of an atom in a V-level configuration exhibiting a chiral coupling with a semi-infinite waveguide, where the atomic transitions are coupled to the modes propagating in opposite directions. The atom is coherently driven by lasers and by its own quantum feedback. In the Markovian limit, i.e. when the retardation effects of the quantum feedback are neglected, we have shown that the dynamics can lead to a situation where the atom is in a pure steady-state and no photons are emitted in the waveguide. The properties of this steady-state crucially depend on the phase acquired by the quantum feedback. 
 
 If this phase is a multiple of $2\pi$, the atom mimics the dynamics of a pair of laser-driven two-level atoms coupled to a unidirectional waveguide in a cascaded setup \cite{1367-2630-14-6-063014}. In the steady-state of the dissipative dynamics, these two-level atoms form a pure EPR-entangled pair which effectively decouples from the waveguide, and the excited fraction of this steady-state grows linearly with the driving intensity. The current effort in the development of chiral atom-waveguide couplings in different platforms \cite{Mitsch:2014aa,Sollner:2015aa} should allow the observation of these atomic pairs in the future. In the meantime, we have shown that a single atom can experience analogous physics which could be achieved with current state-of-the-art technology. 
 
 If the phase of the quantum feedback is different than $2\pi$, we have shown that the pure steady-state of the atom exhibits very different properties, where the driving strength is dependent on the feedback phase and the excited fraction of the steady-state is now a constant. We have also extended our study to non-Markovian regimes, where the retardation effects in the quantum feedback become important, and we have shown that this delay induces a shift in the feedback phase.

\section*{Acknowledgements}
 
The work at Innsbruck is supported by the ERC Synergy Grant UQUAM, the Austrian Science Fund through SFB FOQUS, the EU FET Proactive Initiative SIQS and RYQS, and  the Army Research Laboratory Center for Distributed Quantum Information via the project SciNet.
H.~P. was supported by the NSF through a grant for the Institute for Theoretical Atomic, Molecular, and Optical Physics at Harvard University and the Smithsonian Astrophysical Observatory.
 \appendix
 \section{Adiabatic elimination of the cavity modes}
 \label{appendixcavity}
 We consider the setup of Sec.\,\ref{sec:couplcavi} and we derive the master equation for the atom in the regime $g\ll\kappa$ starting from Eq.\,\eqref{eq:masteqrhoca}. The projection operator for the adiabatic elimination of the cavity modes is \be \mathcal P\rho\equiv \text{Tr}_c(\rho)\otimes \ket{0}\bra{0},\ee where $\text{Tr}_c$ denotes the trace over the cavity modes and $\ket{0}$ is the vacuum state on the cavity subspace. A second order perturbation expansion of the master equation projected on the $\mathcal P$ space provides an equation for the atomic density matrix $\rho_a\equiv\text{Tr}_c(\rho)$ \cite{gardinerzoller_quantumnoise}  
\begin{eqnarray} \label{eq:redmaster}\frac{d\rho_a}{dt}&=&-i[H_a,\rho_a]+\mathcal L'\rho\\ \nonumber&+& \underbrace{\text{Tr}_c\Big(\mathcal P\big(\mathcal L_\text{int}(-\mathcal L_\text{cav})^{-1}\mathcal L_\text{int} (\rho_a\otimes \ket{0}\bra{0})\big)\Big)}_{\mathcal L_\text{eff}\rho_a}. \end{eqnarray}
Using the fact that $\lim_{t\to\infty}e^{\mathcal L_\text{cav} t} = \ket{0}\bra{0}$ (the unique steady-state of $\mathcal L_\text{cav}$ is the vacuum) and $\text{Tr}_c\big(\mathcal L_\text{int} (\rho_a\otimes\ket{0}\bra{0})\big)=0$, the third term of the right-hand-side can be rewritten 

\be \label{eqleff2} \mathcal L_\text{eff}\rho_a= \int_0^\infty dt\ \text{Tr}_c\Big(\mathcal L_\text{int} e^{\mathcal L_\text{cav} t}\mathcal L_\text{int} (\rho_a\otimes  \ket{0}\bra{0})\Big). \ee
We now move to a picture where the operators evolve with $\mathcal L_\text{cav}$. For two operators $O_1$ and $O_2$ acting on the cavity subspace, we will write $\bra{0}O_1(t)O_2(0)\ket{0}$ to denote the vacuum correlations $\text{Tr}_c\big(O_1e^{\mathcal L_\text{cav} t}(O_2  \ket{0}\bra{0})\big)$. Using this notation and the explicit form of $\mathcal L_\text{int}$ from Eq.\,\eqref{eq:Lidef}, the expression from Eq.\,\eqref{eqleff2} can be rewritten \begin{equation}  \label{eq:corrval}\begin{aligned}\mathcal L_\text{eff}\rho_a=\sum_{i,k\in\{S,T\}}g^2\int_0^\infty dt&\bra{0}{a_i(t)a_k^\dagger(0)}\ket{0}[\sigma_k^-\rho_a,\sigma^+_i]\\
+&\bra{0}{a_k(0)a_i^\dagger(t)}\ket{0}[\sigma_i^-,\rho_a\sigma^+_k]. \end{aligned}\end{equation}
We thus have to obtain the expressions for $\bra{0}{a_i(t)a_k^\dagger(0)}\ket{0}$ and $\bra{0}{a_k(0)a_i^\dagger(t)}\ket{0}$. Using the expression of Eq.\,\eqref{eq:defLB} for the Liouvillian, we first solve the equations of motion for $\average{a_T(t)}$ and $\average{a_S(t)}$ for an arbitrary density matrix $\rho$. These equations are obtained by noting that $\frac{d}{dt}\average{a(t)}=\frac{d}{dt}\text{Tr}(a\rho(t))=\text{Tr}(a\mathcal L_\text{cav}\rho).$ Using the expression of $\mathcal L_\text{cav}$ from Eq.\,\eqref{eq:defLB}, we get
\begin{eqnarray} \frac{d}{dt}\average{a_T(t)}&=&-\big(\kappa+\frac{\kappa'}2\big)\average{a_T(t)} -\frac\kappa2 \average{a_S(t)}\\
 \frac{d}{dt}\average{a_S(t)}&=&\frac{\kappa}2\average{a_T(t)} -\frac{\kappa'}2 \average{a_S(t)},\end{eqnarray}
whose solution reads \begin{eqnarray} \label{eq:laky}\average{a_T(t)}=&&e^{-(\kappa+\kappa') t/2}\average{a_T(0)} \\ \nonumber&&- \frac{\kappa t}{2} e^{-(\kappa+\kappa') t/2}(\average{a_T(0)}+\average{a_S(0)})
 \\  \average{a_S(t)}=&&e^{-(\kappa+\kappa') t/2}\average{a_S(0)} \\&&+ \nonumber\frac{\kappa t}{2} e^{-(\kappa+\kappa') t/2}(\average{a_T(0)}+\average{a_S(0)}).  \end{eqnarray}

We now apply the quantum regression theorem \cite{gardinerzoller_quantumnoise} to obtain the vacuum correlations $\bra{0}a_i(t)a_k^\dagger(0)\ket{0}$. For example, by choosing $\rho=a^\dagger_S\ket{0}\bra{0}$, Eq.\,\eqref{eq:laky} provides \be \bra{0}a_T(t)a^\dagger_S(0)\ket{0}=-\frac{\kappa t}2e^{-(\kappa+\kappa')t/2}.\ee The other terms are similarly obtained, and the integral in Eq.\,\eqref{eq:corrval} can be performed, which provides the effective Liouvillian
 \be  \begin{aligned}\mathcal L_\text{eff}\rho_a=&2\gamma\mathcal{D}[\sigma_S^-]\rho_a +\frac{\gamma}2[\sigma_T^+\sigma_S^--\sigma_S^+\sigma_T^-,\rho_a]\\&+ \gamma\frac{\kappa'}{\kappa}\big(\mathcal D[\sigma_1^-]\rho_a+\mathcal D[\sigma_2^-]\rho_a\big),\end{aligned}\ee
where we define $\gamma = (2g)^2\kappa/(\kappa+\kappa')^2$. In the limit $\kappa'\to0$, this Liouvillian describes the coupling of the atom to the guided modes (Eq.\,\eqref{eq:vsystemmasterequation}), if we exchange the labels of the states $\ket{T}=(\ket{e_1}+e^{i\Delta\phi}\ket{e_2})/\sqrt{2}$ and $\ket{S}=(\ket{e_1}-e^{i\Delta\phi}\ket{e_2})/\sqrt{2}$. This can be done by redefining the phase $\Delta\phi$ with an additional $\pi$ shift, in which case the master equation becomes Eq.\,\eqref{eqcavityadiab}.

 \section{Matrix-product state algorithm in the non-Markovian regime}
 \label{appendixb}
 \subsubsection{Quantum Stochastic Schr\"odinger Equation}
We provide here a description of the numerical method developed in \cite{PhysRevLett.116.093601} that we use to study our feedback system in non-Markovian regimes. Let us start from the interaction picture Hamiltonian given by Eq.\,\eqref{eq:hsysdef1} and Eq.\,\eqref{eq:hintdef} and define the Fourier transform operators \be b(t)=\frac1{\sqrt{2\pi}}\int_{\bar\om-\theta}^{\bar\om+\theta}d\omega\,b_\omega e^{i(\bar\omega-\omega )t}.\ee Their commutation relations  can be approximated by a Dirac delta function $[b(t),b^\dagger(t')]\approx \delta(t-t')$ on timescales much larger than the photon correlation time $1/\theta$. The interaction Hamiltonian (Eq.\,\eqref{eq:hintdef}) becomes \be \label{eq:Hintnoise}H_\text{int}(t)=i \sqrt{\gamma}\big(b^\dagger(t)\sigma_1^- + b^\dagger(t-\tau)e^{i\Delta\phi}\sigma_2^- -\text{H.C.} \big),\ee and provides a Quantum Stochastic Schr\"odinger Equation \cite{gardinerzoller_quantumnoise} $\frac{d\ket{\Psi}}{dt}=(H_a+H_\text{int}(t))\ket{\Psi}$ for the system comprising the atom and the waveguide.
 
 \subsubsection{Time discretization}
We discretize time into time-steps of length $\Delta t$ which we take much smaller than the relevant time-scales $1/\gamma, 1/\Omega, 1/|\delta_i|$, but much larger than the photon correlation time $1/\theta$. For a given $\Delta t$ we define the quantum noise increments \be \label{eq:DBdef}\Delta B_k= \int_{t_k}^{t_{k+1}} dt\ b(t)\ee where $t_{k+1}=t_k+\Delta t$, and $[\Delta B_k,\Delta B_{k'}^\dagger]\approx\Delta t\ \delta_{k,k'}$. In this stroboscopic view, the photons are separated into discrete time-bins, which consists of a bosonic Fock space with the corresponding annihilation operators given by the operator $\Delta B_k/\sqrt{\Delta t}$. The Fock basis for each time-bin $k$ is denoted $\{\ket{i_k}, i_k=1,2,...\}$, where $\ket{i_k}=\frac{(\Delta B^\dagger/\sqrt{\Delta t})^{i_k}}{\sqrt{i_k!}} \ket{\text{vac}}$ and $i_k$ is interpreted as the number of photons in the time-bin. We denote the state of the system consisting of the atom and all the time-bins as $\ket{\Psi(t)}$. 

The evolution between two successive discrete times $t_k$ and $t_{k+1}$ is given by a unitary operator $U_k$, such that $\ket{\Psi(t_{k+1})}=U_k\ket{\Psi(t_k)}$. Using the Hamiltonian from Eq.\,\eqref{eq:Hintnoise}, this operator reads \be\begin{aligned}\label{eqdefUk} U_k = \mathcal{T}\text{exp}\Big(-i H_a\Delta t +& \sqrt{\gamma}\big(\Delta B^\dagger_k\sigma_1^- \\ &+ \Delta B^\dagger_{k-m}e^{i\Delta\phi}\sigma_2^- -\text{H.C.} \big) \Big),\end{aligned}\ee
where $\mathcal{T}$ denotes the time-ordering of the $b(t)$ operators appearing if one replaces the $\Delta B$ operators by their definition (Eq.\,\eqref{eq:DBdef}). To first order in $\Delta t$, we will neglect this time-ordering. During each time-step $k$, we see that the atom interacts only with the time-bins $k$ and $k-m$, where we have defined $m=\lfloor{\tau/\Delta t}\rfloor$. The first one physically corresponds to the photons emitted towards the mirror, whereas the second one corresponds to the delayed interaction with the feedback photons. 

We assume that the initial state is of the form $\ket{\Psi(t=0)} = \ket{\psi_a}\otimes_{p=1}^\infty\ket{\phi_p}$, where $\ket{\psi_a}$ denotes the initial state of the atom and $\ket{\phi_p}$ the initial state of the time-bin $p$ (namely the vacuum state in our case), hence the system is initially fully disentangled. After an evolution up to time $t_k$, the entanglement grows and in general the system is of the form $\ket{\Psi(t_k)} = \ket{\psi(t_k)}\otimes_{p=k}^\infty\ket{\phi_p}$, where $\ket{\psi(t_k)}$ denotes the state of the system comprising the atom and the time-bins up to $p=k-1$. We work in the basis $\ket{i_a, i_{k-1}, i_{k-2}, ..., i_1}$, where $i_a\in\{g,e_1,e_2\}$ labels the atomic states. On this basis, $\ket{\psi(t_k)}$ is decomposed as \be \ket{\psi(t_k)}=\sum_{i_a, \{i_p\}}\psi_{i_a,i_{k-1},i_{k-2},...,i_1}\ket{i_a, i_{k-1}, i_{k-2}, ..., i_1}. \ee

\subsubsection{Matrix-product state algorithm}

 The matrix-product state (MPS) \cite{RevModPhys.77.259} approach consists in writing the amplitude as the trace of a product of matrices \begin{equation}\label{eq:tensorform} \psi_{i_a,i_{k-1},i_{k-2},...} = \text{Tr}\big( A[a]^{i_a} A[k-1]^{i_{k-1}}A[k-2]^{i_{k-2}} ... \big), \end{equation}
 where each $A[p]^{i_p}$ is a matrix of finite dimensions $D_p \times D_{p-1}$. The bond dimension $D_p$  represents the entanglement between the different components of the system, more precisely between the two parties formed by a bipartite cut of the time-bins between bins $p$ and $p+1$. The objects $A[p]$ are thus tensors with 2 bond indices encoding the entanglement, and 1 physical index $i_p$. In our case we use open boundary conditions, meaning that $D_a=D_0=1$ at each time $t_k$. Note that by setting a boundary $D_\text{max}$ for the bond dimensions, the numerical cost is bounded by $NdD_\text{max}^2 $ for $N$ bins with a physical dimension $d$. This can be much lower than the usual exponential complexity $d^N$. As discussed in Ref.\,\cite{PhysRevLett.116.093601}, $D_\text{max}$ needs to increases exponentially with the delay $\gamma \tau$ due to long-range correlations between the bins which increase the entanglement entropy. However, given a fixed $\gamma \tau$, the entropy remains constant once the system reaches the steady-state, meaning that the numerical cost increases only linearly with time.
 
 The algorithm then consists in updating the tensors at each time-step in the following way. We first extend the definition of the system by one time-bin. Formally this amounts to writing \begin{equation} \psi_{i_a,i_{k},i_{k-1},...} = \text{Tr}\big( A[k]^{i_{k}}A[a]^{i_a} A[k-1]^{i_{k-1}} ... \big)\label{eq:newa} \end{equation} where $A[k]^{i_{k}}=\delta_{i_{k},0}$ is a $1\times 1$ matrix, as this new time-bin is still completely disentangled from the rest of the system. The Kronecker delta signifies that the state of the time-bin is the vacuum state. The next step is to apply the unitary evolution from Eq.\,\eqref{eqdefUk}. $U_k$ can be seen as a tensor with 6 physical indices, 3 of which are to be contracted with the physical indices of the MPS tensors $A[a], A[k]$ and $A[k-m]$ to obtain the evolved state. Notice that $A[a]$ and $A[k]$ are successive tensors in the state representation (Eq.\,\eqref{eq:newa}), which corresponds to a short-range interaction between the tensors. On the other hand, the interaction with $A[k-m]$ is more involved, as it implies long-range interactions, hence all the bins between $A[k]$ and $A[k-m]$ must be updated in order to account for the entanglement increase arising from this interaction. Various methods for dealing with these interactions exist \cite{Koffel:2012fv,Haegeman:2014wb,Zaletel:2015jx}. Our algorithm is described in details in the Supplemental Material of \cite{PhysRevLett.116.093601} and employs a method proposed in \cite{Schachenmayer:2010ia,Banuls:2006ir}. It consists in exchanging recursively $m-1$ times the state of the $(k-m)$-th bin with that of the $(k-m+1)$-th bin in order to obtain an MPS description of the state where all three interacting tensors are successive tensors. We then merge these three tensors into a tensor with 3 physical indices $i_k, i_a, i_{k-m}$, and locally apply the unitary tensor $U_k$ by contracting these indices. The resulting tensor can finally be brought back into single bin tensors by applying singular value decompositions (SVD), and by exchanging the positions of $A[k]$ and $A[a]$, the system is put back in the form of Eq.\,\eqref{eq:tensorform}, with $k\to k+1$.

\bibliography{Bibliography,HannesBib}
\end{document}